\documentclass[aps,prd,groupedaddress,nofootinbib,preprintnumbers]{revtex4}
\pdfoutput = 1
\usepackage{graphicx}
\usepackage{latexsym}
\usepackage{amsfonts}
\usepackage{amssymb}
\usepackage{amsmath}
\usepackage{slashed}
\usepackage{tabularx,url,color}
\usepackage{hyperref}
\usepackage{multirow}

\newcommand{\gev}{{\ensuremath \text{GeV}}}

\newcommand{\fb}{{\ensuremath \text{fb}}}

\newcommand{\xpass}{${\mathfrak t_w}$}
\newcommand{\xfail}{${\mathfrak t_{-}}$}

\def\ie{{\sl i.e.}~}

\preprint{KCL-PH-TH/2013-{\bf 36}}
\begin{document}

\title{Buckets of Higgs and Tops}

\author{Matthew R.~Buckley$^{1,2}$, Tilman Plehn$^3$, Torben Schell$^3$, and Michihisa Takeuchi$^4$}

\affiliation{$^1$Center for Particle Astrophysics, Fermi National Accelerator Laboratory, Batavia, IL, U.S.A.}
\affiliation{$^2$Department of Physics and Astronomy, Rutgers University, Piscataway, NJ, U.S.A.}
\affiliation{$^3$Institut f\"ur Theoretische Physik, Universit\"at Heidelberg, Germany}
\affiliation{$^4$Theoretical Particle Physics and Cosmology Group, Department of Physics, King's College London, London WC2R 2LS, UK}

\date{\today}

\begin{abstract}
  We show that associated production of a Higgs with a top pair can be
  observed in purely hadronic decays. Reconstructing the top quarks in
  the form of jet buckets allows us to control QCD backgrounds as well as
  signal combinatorics. The background can be measured from side bands
  in the reconstructed Higgs mass. We back up our claims with a
  detailed study of the QCD event simulation, both for the signal and
  for the backgrounds.
\end{abstract}

\maketitle

\tableofcontents

\newpage

\section{Introduction}
\label{sec:intro}

Measuring the couplings of the recently discovered Higgs
boson~\cite{higgs_theo,higgs_ex} to the Standard Model fermions is a
critical part of the investigation of the electroweak symmetry
breaking mechanism at the LHC~\cite{sfitter}.  The Standard Model
coupling to the top quark is expected to be of order unity, making it
a prime target for studying the effects of many different new physics models in
and beyond the Higgs sector~\cite{bsm_review}.  Together with the
Higgs self coupling it dominates the extrapolation of weak--scale
Higgs physics to more fundamental energy scales~\cite{lecture}.
Measuring this parameter will uniquely probe extensions of the Higgs
sector at the weak scale~\cite{sfitter} as well as at high scales.

However, with a production cross section of only ${\cal O}(500~\fb)$
at the 13~TeV LHC, measurements based on the $t\bar{t}H$ channel
are extremely difficult. Search strategies in the fully leptonic and
semi-leptonic decay channels for the top have been suggested in
combination with Higgs decays to $b\bar{b}$~\cite{tth_bb},
$\tau\tau$~\cite{tth_tautau}, and $W^+W^-$~\cite{tth_ww}. These are
challenging through a combination of combinatoric backgrounds,
missing control regions, large QCD uncertainties on the
backgrounds, and low rate. Typical luminosities required for a
$5\sigma$ signal might well be in the 100~fb$^{-1}$ range for 13~TeV
collider energy, with a signal--to--background ratio well below 1:1.

In this paper we provide a feasibility study for the fully hadronic
channel of $t\bar{t}H$ production, {\it i.e.}~a final state consisting
of four $b$-jets plus up to four un-tagged jets.  This channel has
not been studied yet. In fact, there exist only a few
analyses targeting Higgs or new physics searches in purely hadronic
channels without missing energy or leptons, most notably some recent
top pair resonance~\cite{tt_resonance,buckets} and sgluon
searches~\cite{steffen}.  However, the fully hadronic decay channel of $t\bar{t}H$ has two
possible advantages over the leptonic decay modes. First, hadronic
decays of both the tops and the Higgs have the highest branching
ratios of any decay mode. Second, without neutrinos and their missing
momenta, a full reconstruction of the $t\bar{t}H$ final state is
possible, which allows for discrimination of signal and background and
provides the best testing ground in the presence of possible experimental
anomalies. In addition, this eight--parton final state has the highest
jet multiplicity of any widely-considered Standard Model process at
the LHC. Demonstrating our ability to understand and use such events
is an important benchmark in our study of Standard Model physics at the
LHC.\bigskip

We separate the signal events from the large QCD background via a
two-step process. For four $b$-jet events, we first apply
global acceptance cuts, giving us an enriched sample of signal events. We then
reconstruct the tops using the ``bucket algorithm''~\cite{buckets},
which closes the gap between kinematic top reconstruction at threshold
and proper top taggers~\cite{seymour,tagger_review} by targeting
slightly boosted top quarks, with
\begin{equation}
p_{T,t} \sim 100 - 300~\gev \; .
\end{equation}
The algorithm identifies the two top quarks in the event by permuting
over jet assignments to three buckets, minimizing a distance metric on
two of those buckets between the invariant mass of the jets in the
buckets and the mass of the top. The remaining event contains two
$b$-jets, which allow us to reconstruct the Higgs decay with a
probability above 60\%.

It should be noted that the results in this paper deliberately only
rely on a simple cut--and--count method. It allows us to identify many
opportunities for data--driven side band calibration of the
backgrounds, which is crucial to such high--multiplicity searches.  As
the top-reconstruction technique provides a good approximation of the
momenta of all the particles in the original event, more sophisticated
techniques can be used to improve rejection of background and signal
selection.\bigskip

In the next section, we study the major backgrounds to the $t\bar{t}H$
search, including some global background rejection cuts.  The bucket
algorithm is introduced in Section~\ref{sec:buckets}, where we also
give a detailed estimate of the analysis performance. A detailed
discussion of the QCD backgrounds and their simulation are included in
Appendix~\ref{app:QCD}. Finally, in Appendix~\ref{app:recon} we
show the metrics for the top reconstruction.

\section{Multi-jet backgrounds and global cuts}
\label{sec:multijets}

Our analysis aims to extract the fully hadronic final state of
$t\bar{t}H$ production with the decay $H\to b\bar{b}$. For a Higgs
mass of 125~GeV we assume the Standard Model branching ratio to the
$b\bar{b}$ final state of 57.7\%~\cite{decay,xs_group}. We will
require four $b$-tagged jets in the final state, no leptons, and at
least two un-tagged hard jets. We assume jet-based triggers
for hard multi-jet events, similar to all--hadronic $t\bar{t}$
searches~\cite{tt_resonance}. The main four-$b$ backgrounds, ordered
by relevance, are
\begin{alignat}{7} 
pp \to bb\bar{b}\bar{b} 
\qquad \qquad \qquad pp \to t\bar{t}b\bar{b} 
\qquad \qquad \qquad pp \to tt\bar{t}\bar{t} \; . 
\end{alignat}
The corresponding fake-$b$ channels are strongly suppressed if the
experiments reach a 70\% $b$-tagging efficiency with 1\% mis-tagging
probability for light-flavor and gluon jets.
We estimate the rate of the mis-tagged
$b\bar{b}+$multi-jet background to be contribute to the actual
$bb\bar{b}\bar{b}$ rate at the 10\% level, {\ie}below the quoted
uncertainties in the simulations of the primary background. Similarly, we can ignore
the pure QCD events with four mis-tags.\bigskip

For our event simulation we rely on \textsc{Alpgen}~\cite{alpgen} and
\textsc{Madgraph}~\cite{madgraph}, both with a \textsc{Pythia} parton
shower~\cite{pythia}, as well as on \textsc{Sherpa}~\cite{sherpa}. For
the $t\bar{t}H$ signal our main event sample includes zero and one
hard extra jet merged in the \textsc{Ckkw} scheme~\cite{ckkw} in
\textsc{Sherpa}.  In Appendix~\ref{app:QCD} we compare the
\textsc{Sherpa }results with the \textsc{Madgraph} simulation of
$t\bar{t}H$ plus up to one hard jet merged in the \textsc{Mlm}
scheme~\cite{mlm}. We confirm that the sensitivity to simulation and
QCD issues is minimal. Similarly, for the $t\bar{t}b\bar{b}$
background, our main sample of events is produced by \textsc{Sherpa} and
includes up to one hard QCD jet. A comparison with \textsc{Alpgen}
samples in Appendix~\ref{app:QCD} again shows negligible
dependence on the simulation techniques. We
normalize the merged event samples to the NLO results of 504~fb for
the signal~\cite{tth_nlo,xs_group} and 1037~fb after generator cuts
for the $t\bar{t}b\bar{b}$ background~\cite{ttbb_nlo}.  The
$tt\bar{t}\bar{t}$ background from \textsc{Alpgen} is small compared
to the primary $t\bar{t}b\bar{b}$ and $bb\bar{b}\bar{b}$ backgrounds,
with a cross section of at maximum 5\% of the signal.  As a result, it
does not require an extensive study of the theoretical and simulation
uncertainties, and will not be considered in detail.\bigskip

The critical background for the hadronic $t\bar{t}H$ signal with $H
\to b\bar{b}$ decays is the QCD process $bb\bar{b}\bar{b}+$jets.
Before any selection cuts, its total rate completely overwhelms the
signal, with a cross section of 400~pb estimated by \textsc{Alpgen}
after pre-selection cuts.  However, as any QCD process it is dominated
by soft $b$ and un-tagged jets with an additional enhancement from the
gluon splitting $g \to b\bar{b}$. To extract our signal we will
require four hard, well separated $b$-tagged jets. We 
simulate these background events both in \textsc{Alpgen} and
\textsc{Sherpa} with a hard process of (at least) four $b$-jets.

As we will see in Section~\ref{sec:buckets}, our bucket reconstruction
of two tops and the Higgs will require at least two additional hard
un-tagged jets. Therefore, our central background simulation is
defined by the hard process $bb\bar{b}\bar{b}jj$ plus parton shower
in \textsc{Alpgen}, which results in a cross section of 2128~fb after pre-selection cuts.
To ensure that our analysis is stable with respect to QCD
uncertainties, we also simulate the background with \textsc{Sherpa} as
$bb\bar{b}\bar{b}$ plus zero and one matrix element jet merged
($bb\bar{b}\bar{b}$+0/1j).  The computational expense of two jet
merging is prohibitive here, and so is not included.  However, to have
a measure for the underlying theory uncertainties, we vary the
renormalization and factorization scales in the \textsc{Sherpa}
simulation by a factor 1/2 to 2 around the central scale, to ensure
that our conclusions hold independent of these choices.  We carefully compare
our two background estimates in Appendix~\ref{app:QCD},
providing detailed information on kinematic distributions and the
different simulation tools and hard processes.\footnote{We would like
  to thank the referees and the editor of Ref.~\cite{buckets} for
  strongly supporting this kind of analysis and then allowing us to
  postpone it to this paper.}  There, we test a couple of important
assumptions underlying our analysis.  First, we demonstrate that the
bucket analysis allows only background events with at least two hard
un-tagged jets in our signal region.  For this region the
\textsc{Alpgen} estimate of the $bb\bar{b}\bar{b}jj$ rate is the
appropriate and conservative choice.  In addition, we demonstrate that
our analysis is not too sensitive to describing the second un-tagged jet
by either the hard matrix element (as in the \textsc{Alpgen}) or by
the parton shower (as in \textsc{Sherpa}).  Finally, the full merged
$bb\bar{b}\bar{b}$+jets simulation would allow access to excellent
control regions in the side band of the number of un-tagged jets, once
these kinds of $n_\text{jet}$ distributions are systematically
evaluated by ATLAS and CMS~\cite{jet_scaling}.

Understanding the kinematics of the $bb\bar{b}\bar{b}$ background and
reducing it using global kinematic cuts will be the central topic of
this section. In the next section, we will find that several of these
kinematic cuts can be replaced with the requirement of top
reconstruction, which allows for an increased purity of signal over
background.  In the actual buckets analysis in
Section~\ref{sec:buckets}, we only quote the \textsc{Alpgen} results
for the $bb\bar{b}\bar{b}$ background.\bigskip

Compared to the signal rate, the raw QCD background (primarily
$bb\bar{b}\bar{b}+$jets) is overwhelmingly large, and so we must apply
selection cuts before the top-finding bucket algorithm can be
employed. First, we require all events to have four $b$-tagged
jets. These $b$-jets must be central and widely separated,
to avoid the phase space regions with enhanced $g \to b\bar{b}$
splitting, with
\begin{equation}
p_{T,b} > 40~\gev , 
\qquad |\eta_b| <2.5, 
\qquad \Delta R_{bb} >1.0 
\qquad (4 \times) \; .  
\label{eq:global_cuts}
\end{equation}
In addition, we require at least two hard non-$b$ jets with
\begin{equation}
p_{T,j} > 40~\gev , 
\qquad |\eta_j|< 4.5,
\qquad \Delta R_{jj}> 0.5 
\qquad (2\times) \; . 
\label{eq:global_jets}
\end{equation}
These naive acceptance cuts are very inefficient, for example when compared
 to sub-jet methods. However, the aim of this paper is to show
that the purely hadronic $t\bar{t}H$ process can be studied at the
LHC, so we need to ensure that the pure QCD backgrounds can be
reliably removed. Moreover, four individual $b$-tags cannot be treated
as statistically independent unless we at least assume very widely
separated $b$-jets. This necessitates the harsh cuts in this
proof--of--concept analysis.\bigskip

\begin{table}[t]
\begin{tabular}{l|rrr|r}
\hline
& $t\bar{t}H$ & $t\bar{t}b\bar{b}$ & $ bb\bar{b}\bar{b}jj$ &  $S/B$ \cr
\hline
After acceptance Eqs.\eqref{eq:global_cuts} and \eqref{eq:global_jets}
& 1.197 & 8.363 & 54.420 & 0.019\cr
After global cuts Eq.~\eqref{eq:meff_cuts}
&0.134 & 0.558 & 2.734 & 0.041 \cr
\hline
 \multicolumn{5}{c}{Mass window $m_{bb} = 90 - 130$~GeV} \cr \hline
closest & 0.096  & 0.299 & 1.577 & 0.051 \cr 
hard & 0.017 &  0.031 & 0.226 & 0.065  \cr 
soft & 0.060 &  0.173 & 0.893 & 0.056 \cr 
min & 0.071 &  0.246 & 1.143 & 0.051  \cr 
\hline
\end{tabular}
\caption{Cross section (in fb) of signal and background events after
  successive selection cuts. The $bb\bar{b}\bar{b}jj$ rate is based
  on the \textsc{Alpgen} simulation.  After the full set of cuts from
  Eqs.\eqref{eq:global_cuts}, \eqref{eq:global_jets}, and
  \eqref{eq:meff_cuts}, we show several naive ways of selecting two
  $b$-jets to reconstruct the Higgs mass: the pair closest
  in invariant mass to the Higgs, the two hardest $b$-jets, the two softest $b$-jets, and
  the two $b$-jets with the minimum invariant mass.  We assume a 70\% $b$-tagging efficiency and
  neglect the small mis-tag backgrounds.}
\label{tab:global}
\end{table}

\begin{figure}[b!]
\includegraphics[width=0.32\textwidth]{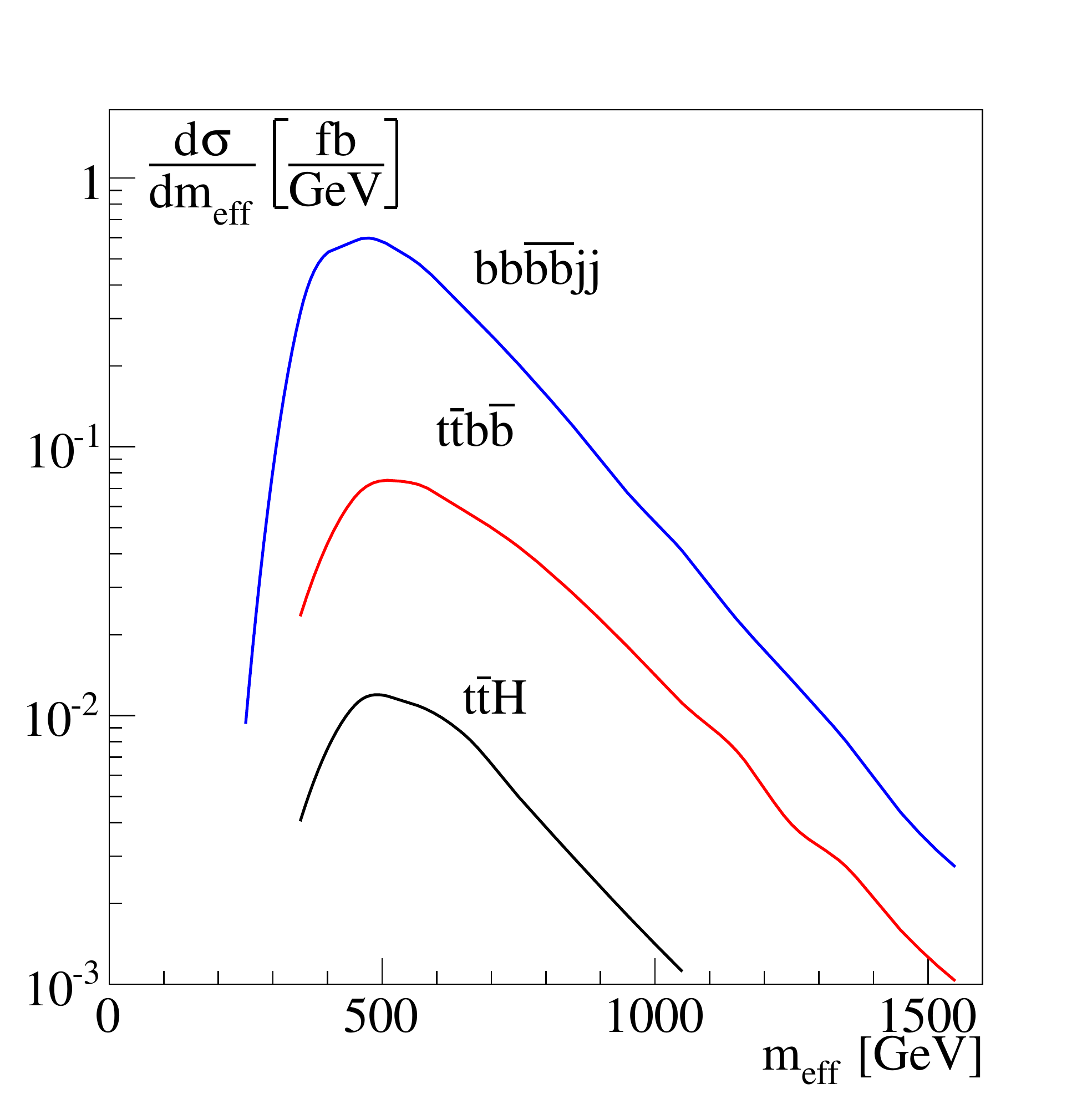}
\includegraphics[width=0.32\textwidth]{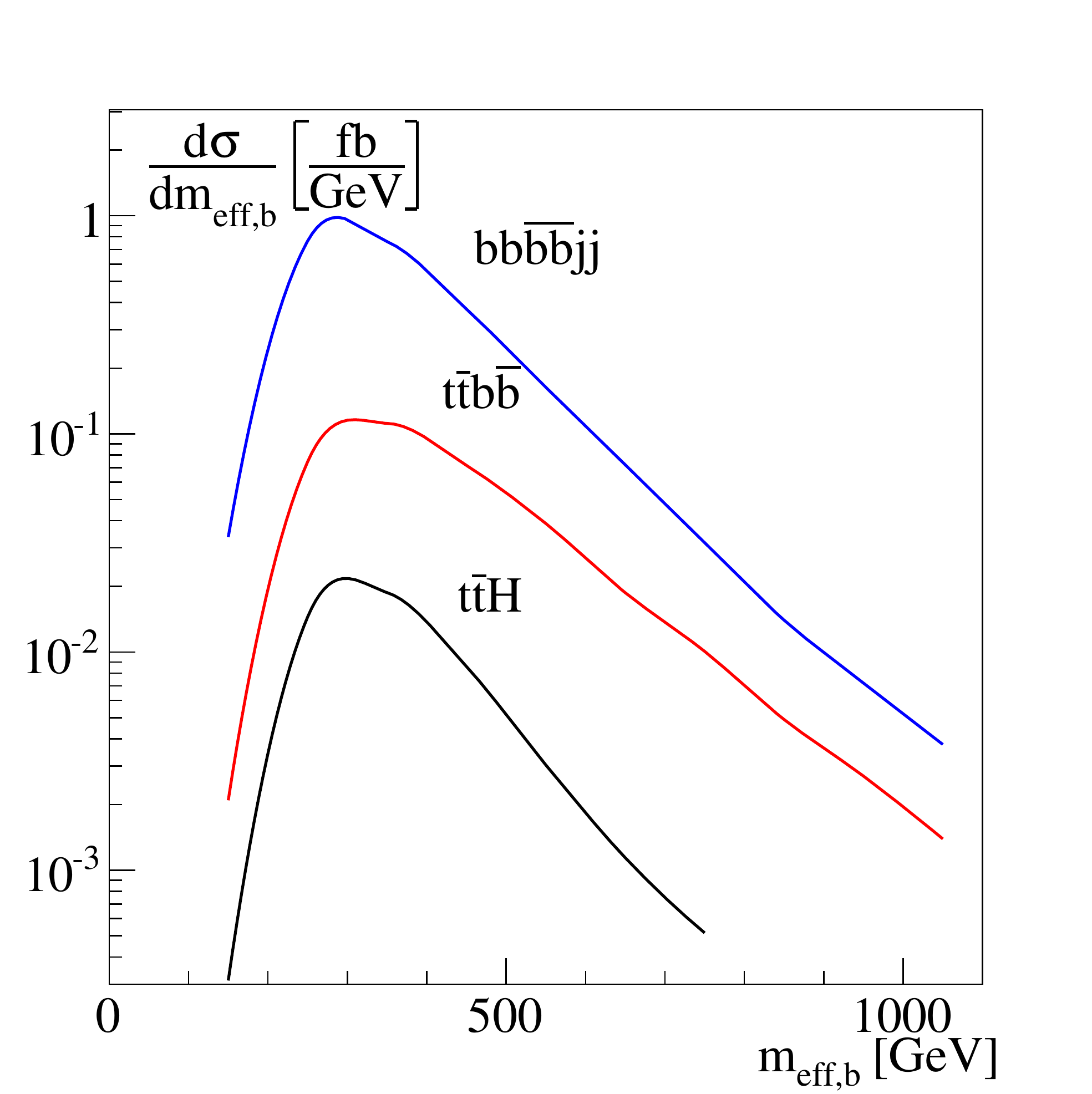}
\includegraphics[width=0.32\textwidth]{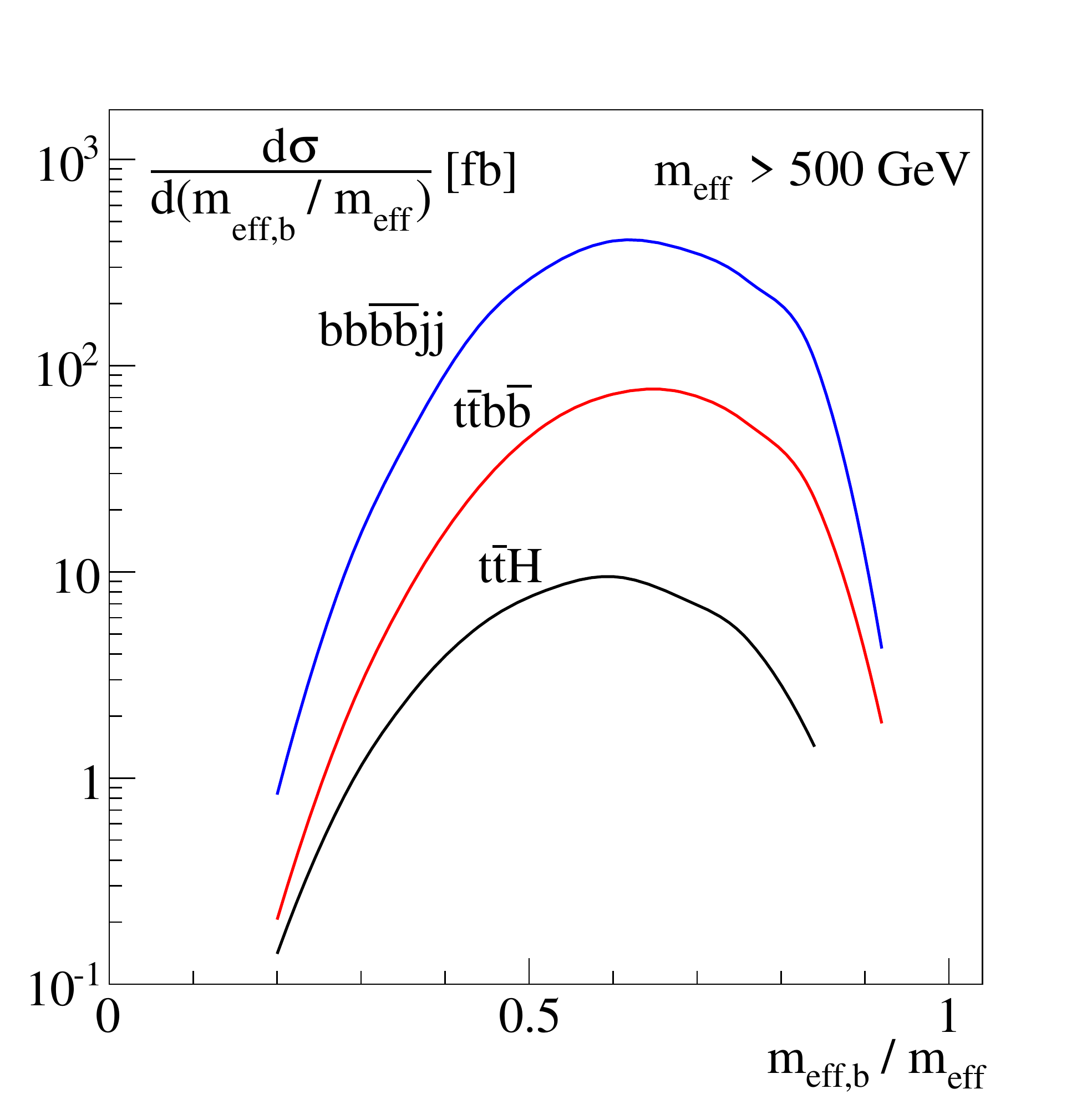}
\caption{Signal and background distributions for the effective mass of
  the entire jet system, the four $b$-tagged jets, and their ratio.
  All jets fulfill Eq.\eqref{eq:global_cuts} and
  Eq.\eqref{eq:global_jets}. We require $m_\text{eff} >
  500$~GeV in the selection cuts of Eq.\eqref{eq:meff_cuts}. For the $bb\bar{b}\bar{b}$ background we show the
  \textsc{Alpgen} result with two additional hard jets plus parton
  shower.}
\label{fig:global}
\end{figure}

In the first line of Table~\ref{tab:global} we show the cross sections
for the signal and two primary backgrounds at the 13~TeV LHC, after
acceptance cuts. The $tt\bar{t}\bar{t}$ contribution is sub-percent
level, and so not shown.  At this stage the $bb\bar{b}\bar{b}$+jets cross section is
still significantly larger than the signal, so an additional set of
cuts is required. Once we introduce the top reconstruction technique,
such cuts are not necessary, but it will be instructive to compare our
later results to simple global cuts.  We consider two global variables: the
effective mass calculated by summing the scalar jet $p_T$ over all jets,
including those with $b$-tags, and its counterpart where the sum runs
only over the four $b$-tagged jets,
\begin{alignat}{7}
m_\text{eff} = \sum_\text{all jets} p_T \; ,
\qquad \qquad  \qquad 
m_{\text{eff}, b} = \sum_\text{four $b$-jets} p_T \; .
\label{eq:def_meff}
\end{alignat}
Both of these observables will be sensitive to the kinematics of the
multi-jet system. In Figure~\ref{fig:global} we show the distributions
for both signal and backgrounds, normalized to the event rates after
the cuts of Eqs.\eqref{eq:global_cuts} and~\eqref{eq:global_jets}. At
this point, the signal--to--background ratio is around 1:50. As we
will discuss in Appendix~\ref{app:QCD}, the fact that the
$m_\text{eff}$ distributions of the signal and the $bb\bar{b}\bar{b}$
backgrounds show a similar behavior is because our \textsc{Alpgen}
simulation requires two hard un-tagged jets. In other words, the
background simulation shown in Figure~\ref{fig:global} anticipates the
fact that we will only be interested in a reliable prediction of those
background events which are kinematically similar to the signal. The
right panel of Figure~\ref{fig:global} shows that after requiring
$m_\text{eff} > 500$~GeV both $m_\text{eff}$ and $m_{\text{eff},b}$
have similar shapes for signal and background.\bigskip

\begin{figure}[t]
\includegraphics[width=0.30 \textwidth]{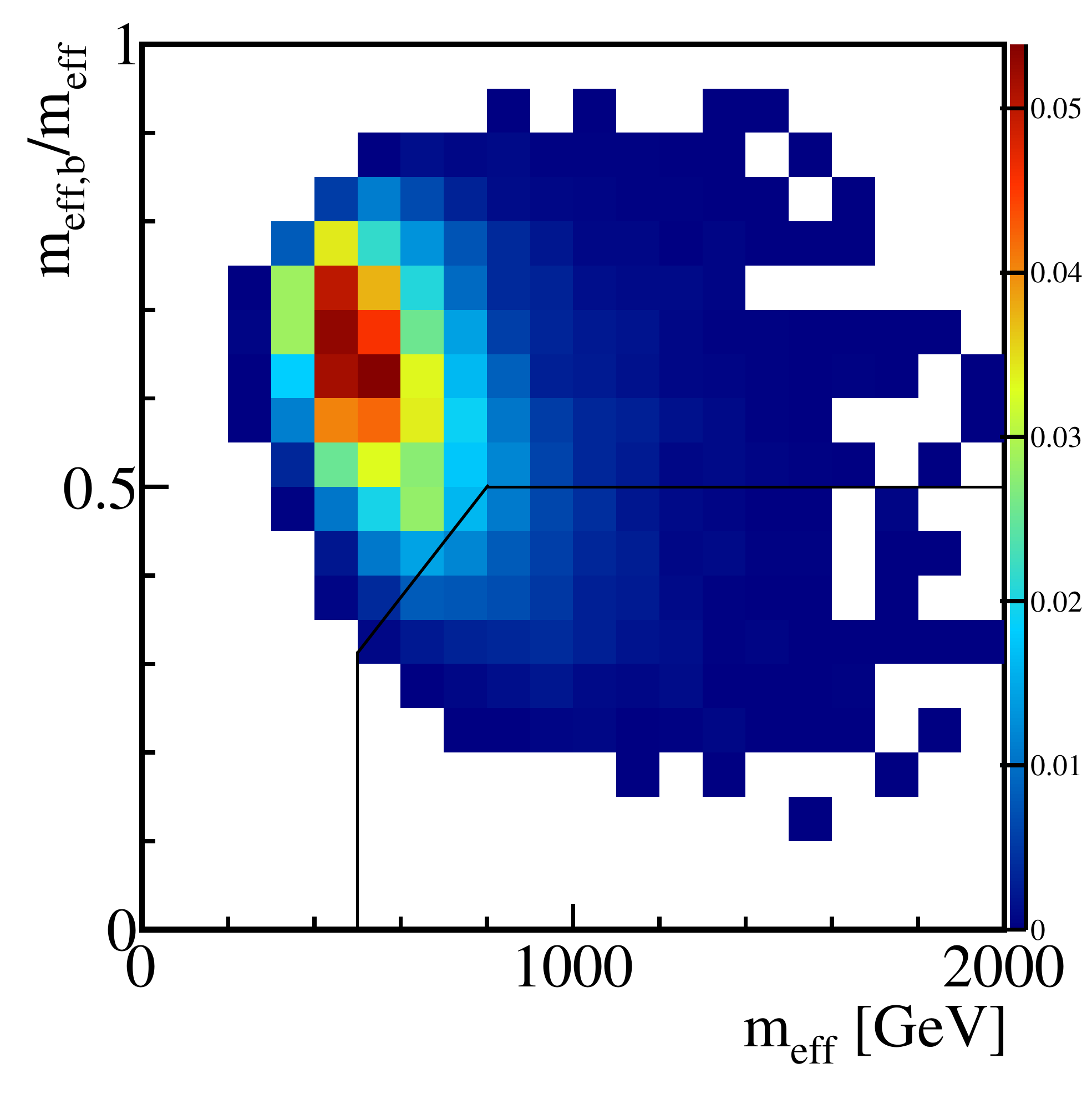}
\hspace*{0.02\textwidth}
\includegraphics[width=0.30 \textwidth]{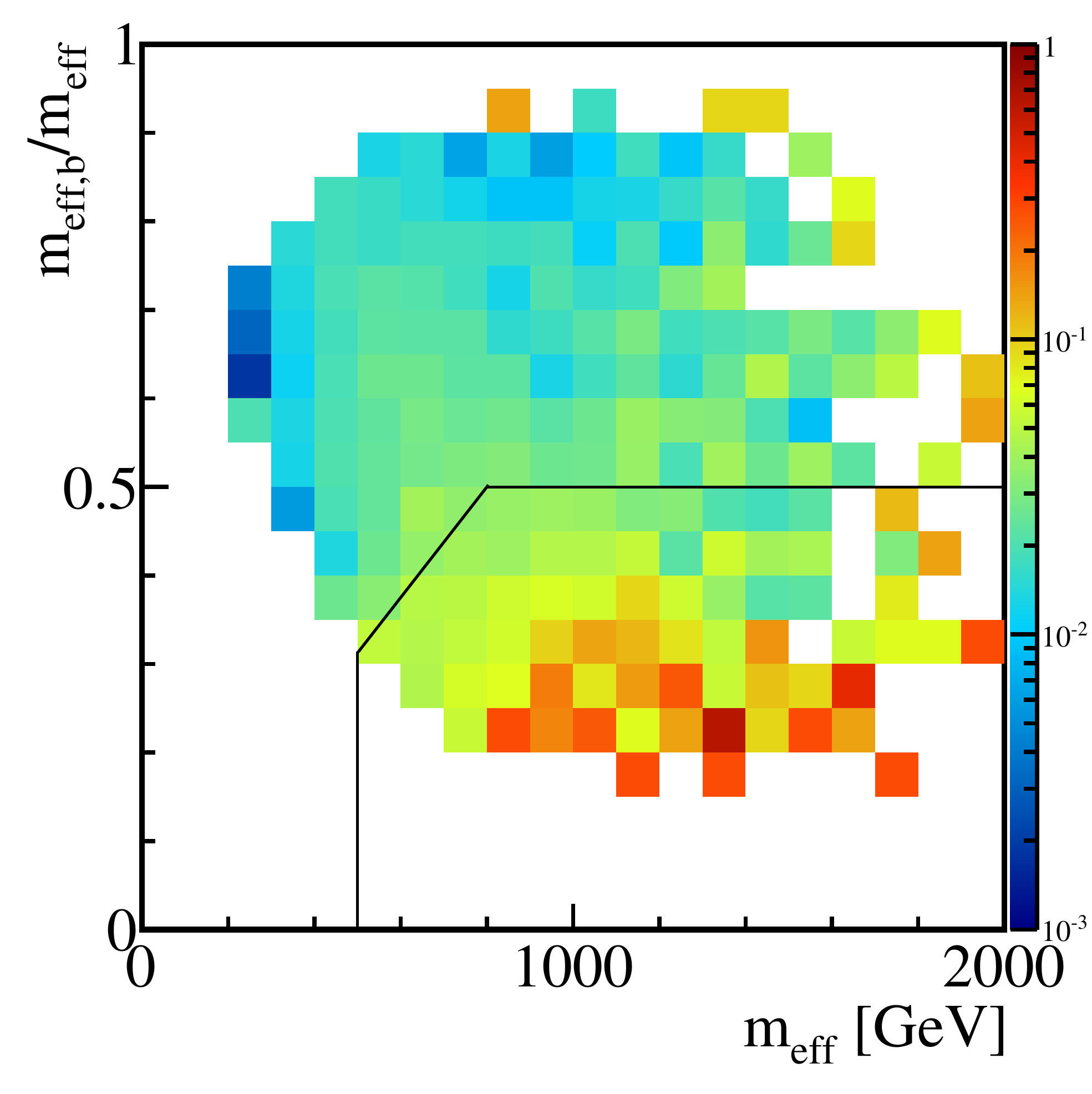}
\hspace*{0.02\textwidth}
\includegraphics[width=0.30 \textwidth]{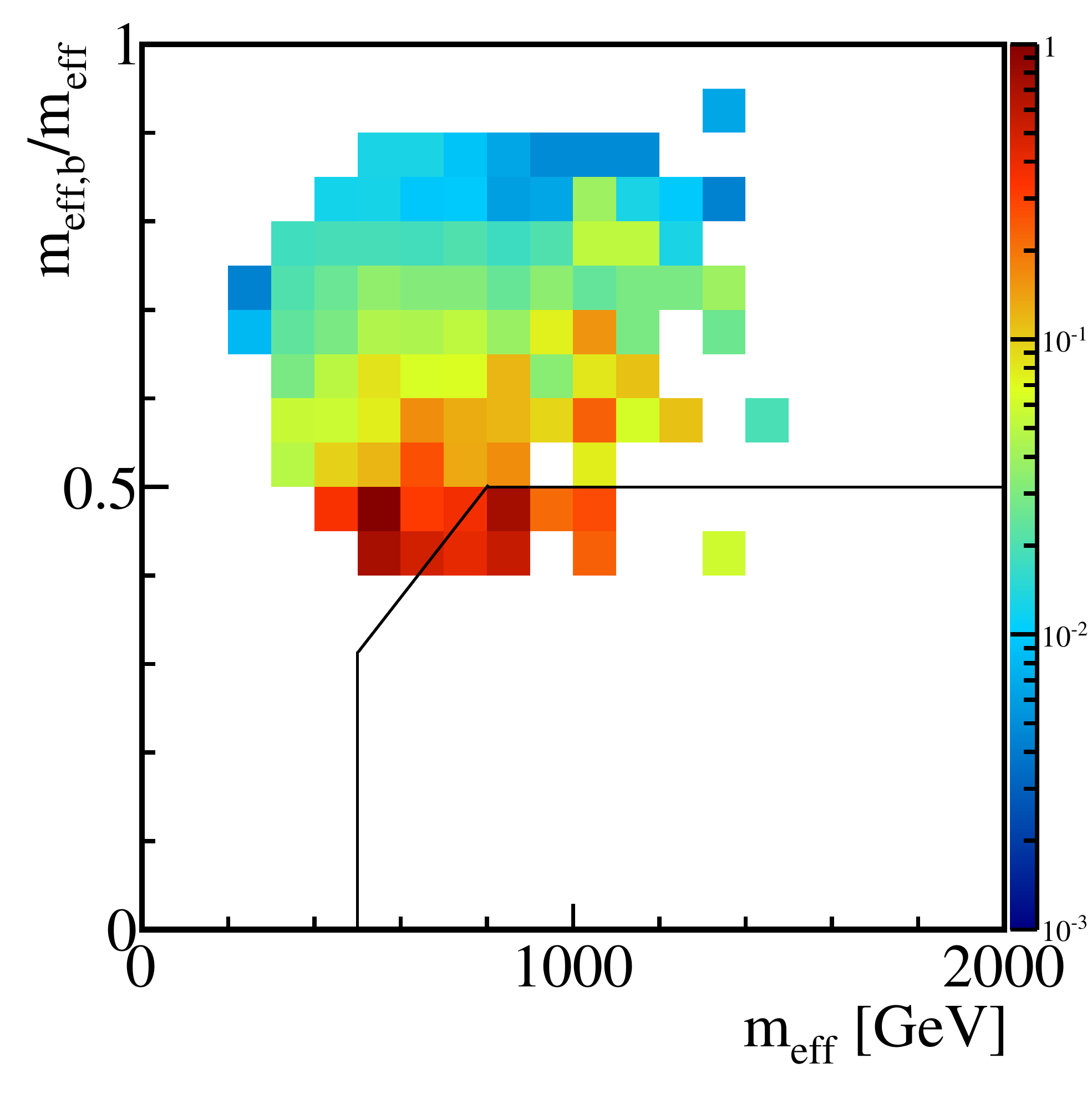}
\caption{Two-dimensional distribution of $m_\text{eff}$
  vs.~$m_{\text{eff},b}/m_\text{eff}$ for the $t\bar{t}H$ signal
  (left), and the ratio of the signal to the $bb\bar{b}\bar{b}jj$
  \textsc{Alpgen} background (center), and the ratio of signal to
  $bb\bar{b}\bar{b}+0/1j$ \textsc{Sherpa} simulation (right).  The lines
  represent the cuts of Eq.\eqref{eq:meff_cuts}.}
\label{fig:meff_bratio_2d}
\end{figure}

It is more efficient to consider the 2-dimensional plane of the two
effective mass variables defined in Eq.\eqref{eq:def_meff}.  In
Figure~\ref{fig:meff_bratio_2d}, we plot these distributions for the
signal and the ratios of signal--to--background against the primary
($bb\bar{b}\bar{b}+$jets) for both \textsc{Alpgen} and \textsc{Sherpa}
simulations. As can be seen, the \textsc{Sherpa} simulation has many
fewer events in the high $m_\text{eff}$ tail compared to the
\textsc{Alpgen} simulation, as expected due to the acceptance cut.
The \textsc{Sherpa} simulation only generates up to one light-flavor or gluon
jet from the hard matrix element, but the acceptance cuts require at
least two hard un-tagged jets per event. As argued in
Appendix~\ref{app:QCD} we use the $bb\bar{b}\bar{b}+$jets sample from
\textsc{Alpgen} for a more conservative background estimate.  In this
proof--of--concept paper, we make the crude requirements that
\begin{equation}
m_\text{eff} > 500~\gev \; ,\qquad \qquad
\frac{m_{\text{eff},b}}{m_\text{eff}} < 0.5  \;, \qquad \qquad 
\frac{m_{\text{eff},b}}{m_\text{eff}} < \frac{m_\text{eff}}{1600~\gev} \; . 
\label{eq:meff_cuts}
\end{equation}
This set of cuts brings the background rate to a manageable level,
without a detailed analysis of the top and Higgs kinematics.  For the
specific $bb\bar{b}\bar{b}jj$ background modeling with
\textsc{Alpgen} we arrive at $S/B \sim 1/25$, as quoted in
Table~\ref{tab:global}.\bigskip

From this point on, we are interested in identifying an excess of
events that contain two $b$-jets which are clearly identified with the
Higgs boson decay. This is complicated by the combinatorial background
of picking the correct two $b$-jets out of the four in the event. First,
we consider naive set of selection criteria for the two $b$-jets which
have to lie in the Higgs mass window in
Table~\ref{tab:global}.  We show that selecting the two $b$-jets
closest in mass to $m_H = 125$~GeV, the two $b$-jets with the softest
$p_T$, the two hardest, and the two $b$-jets with the minimum invariant mass are all
methods that fail to sufficiently increase signal over background.

Clearly, a better solution to the reconstruction of the top and Higgs
decay products and the combinatorics associated with this assignment
is needed. We therefore turn to the bucket
reconstruction~\cite{buckets} to rebuild the two top quarks in the
event, using those events that have passed our initial selection
criteria described by Eqs.\eqref{eq:global_cuts} and
\eqref{eq:global_jets}. This tags the two $b$-jets that come from the
tops with a good degree of accuracy, identifying the Higgs decay
products by exclusion. With this method of identifying the correct
$b$-jets, the global cuts on $m_\text{eff}$ variables do not improve
the $S/B$ ratio, and so we do not continue to apply the cuts of
Eq.\eqref{eq:meff_cuts}. This simple algorithm is not meant to replace
a full experimental likelihood analysis, but it shows that after
simple kinematic cuts a purely hadronic $t\bar{t}H$ analysis can be a
realistic possibility.\bigskip

\section{Top buckets}
\label{sec:buckets}

Following the arguments in the last section and the corresponding
Appendix~\ref{app:QCD} it should be possible to devise an analysis of
the hadronic top and Higgs kinematics to reduce the backgrounds to a
manageable level. Aside from the irreducible $t\bar{t}b\bar{b}$
background we need to extract the signal from the huge multi-jet
$bb\bar{b}\bar{b}$ background.  A more specific analysis of the
multi-jet final state should be able to do better than the already
promising global effective mass cuts in Eq.\eqref{eq:meff_cuts}. The
key concern will be the signal efficiency, because of the limited
$t\bar{t}H$ rate.  For this reason, we choose the bucket
reconstruction~\cite{buckets}, which allows us to keep a larger
fraction of signal events while removing significant parts of the background
phase space identified by the global cuts analysis.  The technical
challenge is tracking the definition of the signal region
and the corresponding background simulation.\bigskip

After applying the jet-level selection cuts in the previous section,
Eqs.\eqref{eq:global_cuts} and \eqref{eq:global_jets}, we have a
sample of events with four $b$-jets and additional extra jets. Of
these four $b$-jets, two are presumed to come from the Higgs decay, and
two from top decays.  Without knowledge of the top decays, various
Higgs reconstruction schemes could be tried. As discussed in the
previous section, one could take the two $b$-jets with the highest or
lowest $p_T$, the combination of $b$-jets with invariant mass that is
closest to 125~GeV, the combination with the minimum invariant
mass, or some other set based on simple jet kinematics.
Taking the combination with the invariant mass closest to that of the
Higgs in particular runs into a combinatorial problem: in both signal
and background, one can often find two $b$-jets with invariant mass
near that of the Higgs without the jets involved having originated
with the Higgs. This shapes the background to look like signal. The
multi-$b$ combinatorics are the reason that the ATLAS
$t\bar{t}H$ search in the early phase of LHC running was largely
abandoned~\cite{cammin}.

We can improve this situation if we find a better way to identify the
$b$-jets that come from the Higgs. We approach this problem by first
identifying the decay products of the tops, using the top bucket
algorithm. The idea behind this algorithm is very simple and
straightforward: we divide all jets in every event into three
buckets. Two of the buckets correspond to the hadronic tops, while the
third bucket consists of all jets in the event that cannot be
associated with a top. In the original formulation of the
algorithm~\cite{buckets} this last bucket was identified with initial
state radiation (ISR). In the current analysis, this ISR bucket will
contain two $b$-jets, which can --- by exclusion --- be identified as
the decay products of the Higgs.\bigskip

We start by seeding each of the two top buckets with a $b$-jet. We
permute over all possible assignments of $b$-jets as top bucket seeds.
We then cycle through every possible assignment of non-$b$-tagged
jets to the three buckets, requiring at least one non-tagged jet in
each of the top buckets. We use the distance metric
\begin{equation}
\Delta_{B_i} = |m_{B_i} - m_t|
\qquad \text{with} \quad 
m_{B_i}^2 = \left(\sum_{k \in B_i} p_k \right)^2 \; , 
\label{eq:delta}
\end{equation}
where $m_t$ is the top mass and the sum runs over all jets (including
the $b$-jet) in the bucket $B_i$.  We select the jet assignment that
minimizes the distance measure $\Delta^2 = \omega
\Delta_{B_1}^2+\Delta_{B_2}^2$, where $\omega > 1$ is a factor chosen
to stabilize the jet grouping. For this analysis, we choose $\omega =
100$, which essentially decouples the second bucket from the
metric. Thus, bucket $B_1$ is the bucket with invariant mass closest
to the top.  After this construction, we have two buckets $B_1$ and
$B_2$, with two or three jets, including the seed $b$-jet. Rarely, we
find a bucket containing four or more jets, in which case we merge to
three jets using the Cambridge/Aachen algorithm~\cite{ca_algo}. 

To remove background events that do not contain real tops, we require
the invariant masses of the two top buckets to lie in the window
\begin{equation}
155~\gev < m_{B_{1,2}} < 200~\gev.
\end{equation}
Next, we require both $B_1$ and $B_2$ buckets to contain a
hadronically decaying $W$ boson candidate. We define a mass ratio cut,
as in the {\sc HEPTopTagger}~\cite{heptop_stops},
\begin{equation}
\left|\frac{m_{k\ell}}{m_{B_i}} - \frac{m_W}{m_t}\right| <0.15
\label{eq:Wcut}
\end{equation}
for at least one combination of the non-$b$-jets (denoted $k$ and
$\ell$) in the bucket $i$. Buckets with only one $b$-tagged and one
non-tagged jet by construction cannot satisfy Eq.\eqref{eq:Wcut}. We
can therefore classify events in one of three categories:
\begin{itemize}
\item (\xpass,\xpass): both top buckets have $W$ candidates as defined by Eq.\eqref{eq:Wcut},
\item (\xpass,\xfail) or (\xfail,\xpass): only the first or second top bucket has a $W$ candidate,
\item (\xfail,\xfail): neither top bucket has a $W$ candidate.
\end{itemize}
The \xpass\, or \xfail\, status is ordered as $(B_1,B_2)$, where
again, $B_1$ is defined as the bucket closest in mass to the
top. Buckets classified as \xpass\, have to include at least three
jets, while \xfail\, buckets can include either three or two
jets.\bigskip

For buckets that fail the criteria of Eq.\eqref{eq:Wcut}, we can still
attempt to reconstruct a top by replacing Eq.\eqref{eq:delta} with an
alternative distance metric,
\begin{equation}
\Delta^{bj}_B = \left\{ \begin{array}{ll} |m_B - 145~\gev|  & \qquad \text{if} \ m_B \le 155~\gev\\\infty & \qquad  \text{else}\\
\end{array}
\right. \; .
\label{eq:Bjwindow}
\end{equation}
We re-assign all $b$-tagged and un-tagged jets in the \xfail~bucket(s), 
combined with the jets in the ISR bucket to new buckets,
irrespective of their original categorization. For this re-assignment
we minimize $\sum_i \Delta^{bj}_{B_i}$. For a top candidate, we
require at least one $b$/jet pair satisfying
\begin{equation}
75~\gev < m_{bj} < 155~\gev.
\end{equation}
This revamped metric is intended to capture top events where the less
energetic jet from $W$ decay was lost in the detector.\bigskip

\begin{table}[b!]
\begin{tabular}{r|rrrr|r}
\hline
& $t\bar{t}H$ && $t\bar{t}b\bar{b}$ &$ bb\bar{b}\bar{b}jj$ & $S/B$ \cr
\hline
After acceptance cuts Eqs.\eqref{eq:global_cuts} and \eqref{eq:global_jets}
& 1.197 && 8.363 &54.420 & 0.019\cr
\hline
2 tops tagged  & 0.894 & (0.184) & 5.882 & 29.356 & 0.025 \cr 
$p_{T,t,1}>100$ GeV &0.709 & (0.158) & 4.868 & 20.838 & 0.028 \cr 
$p_{T,t,1}>200$ GeV &0.289 & (0.080) & 2.189 & 5.194 & 0.039 \cr 
$p_{T,t,1}>300$ GeV &0.089 & (0.028) & 0.724 & 0.917 & 0.054 \cr 
\hline
 \multicolumn{6}{c}{Mass window $m_{bb} = 90 - 130$~GeV} \cr \hline
2 tops tagged  & 0.259 & (0.121) & 0.859 & 5.424 & 0.041 \cr 
$p_{T,t,1}>100$ GeV &0.208 & (0.105) & 0.688 & 3.600 & 0.048 \cr 
$p_{T,t,1}>200$ GeV &0.091 & (0.054) & 0.265 & 0.679 & 0.096 \cr 
$p_{T,t,1}>300$ GeV &0.028 & (0.019) & 0.072 & 0.082 & 0.182 \cr 
\hline
\end{tabular}
\caption{Cross sections (in fb) of events after the acceptance cuts of
  Eqs.\eqref{eq:global_cuts} and \eqref{eq:global_jets} and requiring
  two tops passing the bucket reconstruction. We also require one of
  the reconstructed tops to pass various $p_T$ thresholds, with and
  without requiring the two remaining $b$-jets to have invariant mass
  inside the window 90-130~GeV. Number in parenthesis correspond
  to the events where the reconstructed Higgs lies within  $\Delta R < 0.5$ of the true Higgs.}
\label{tab:simplebb}
\end{table}

\begin{figure}[t]
\includegraphics[width=0.32\textwidth]{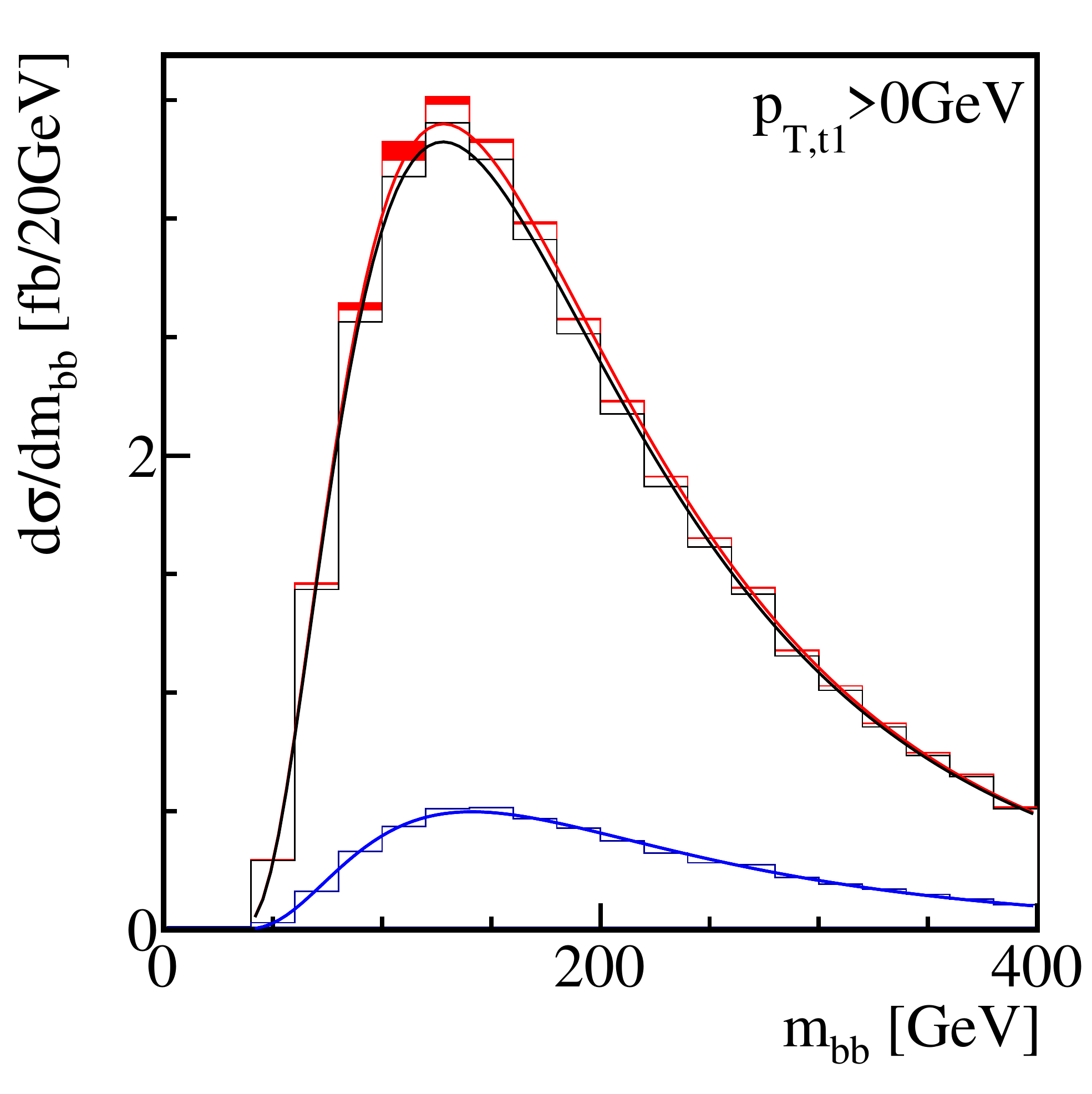}
\includegraphics[width=0.32\textwidth]{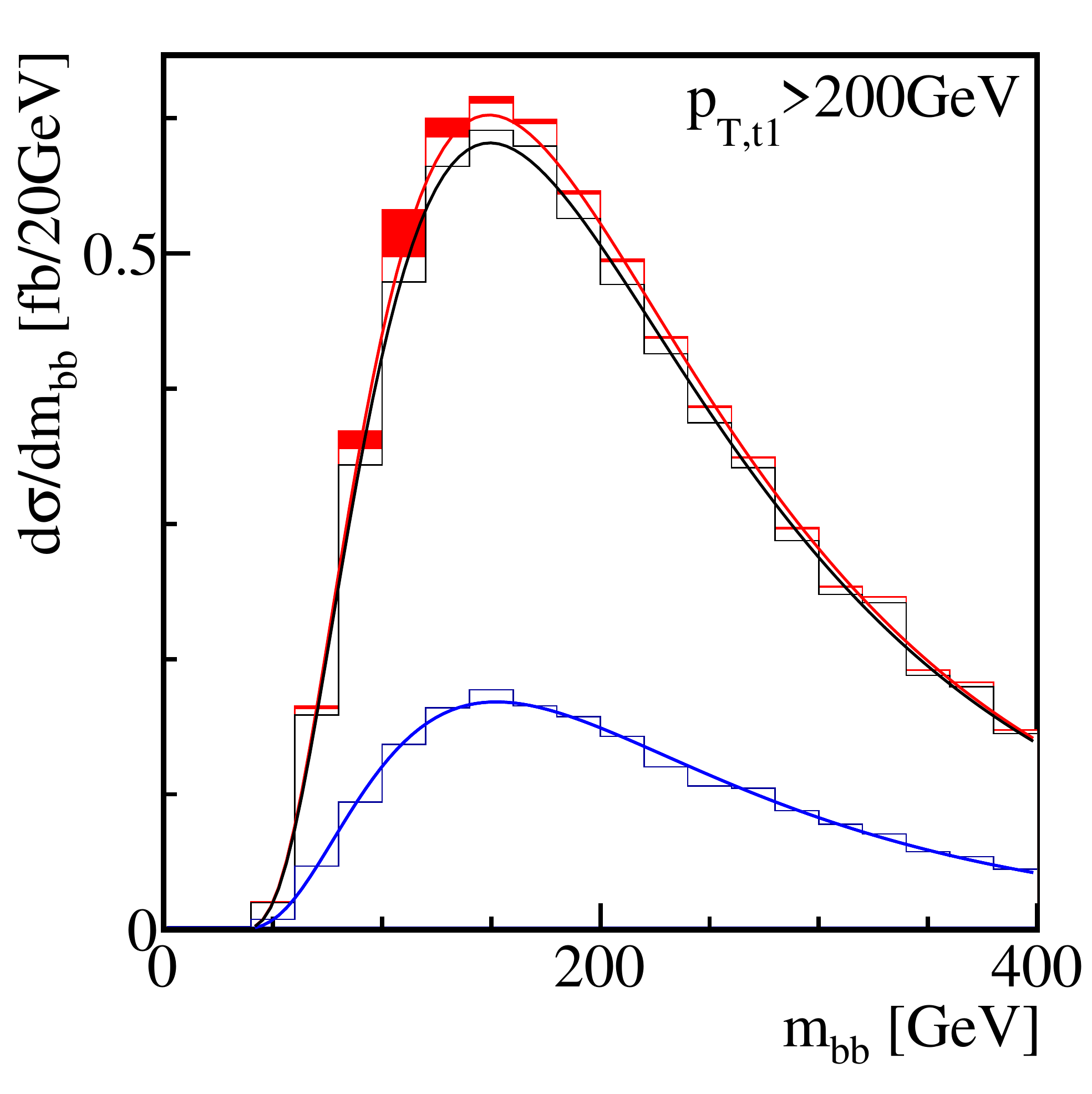}
\includegraphics[width=0.32\textwidth]{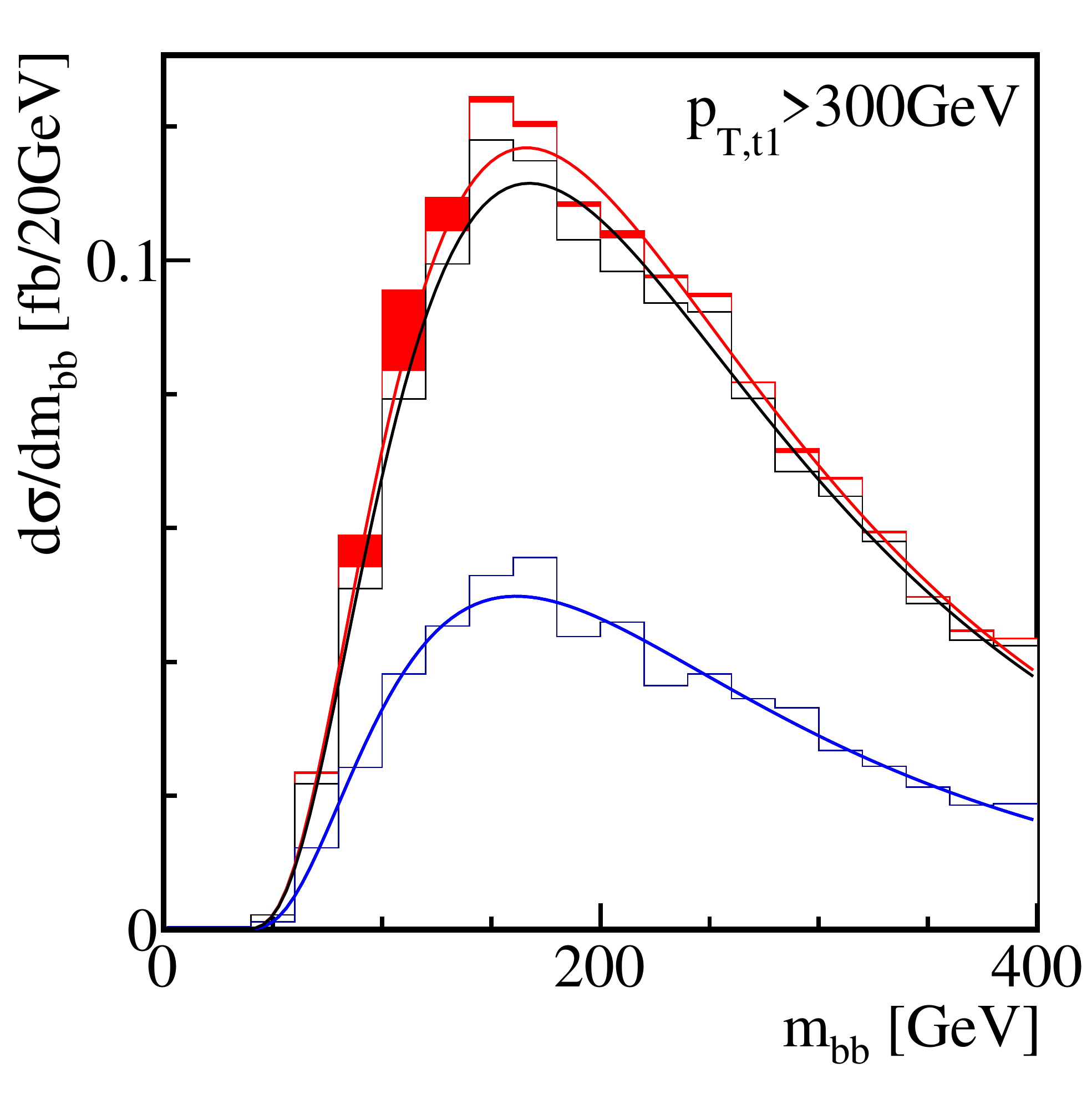}
\caption{Stacked $m_{bb}$ distribution, built from $b$-jets not used
  for top buckets, after top reconstruction and requiring the leading
  top $p_{T,t_1} > 0, 200,$ and 300~GeV (from left).  The two primary
  backgrounds are $t\bar{t}b\bar{b}$ (blue) and $bb \bar{b}\bar{b}$
  (black). The signal distribution of reconstructed tops is shown in
  red. The subset of signal events where the reconstructed Higgs lies
  within $\Delta R < 0.5$ of the true Higgs are displayed as filled-in
  red regions.}
\label{fig:mbb_left}
\end{figure}

The rate of $t\bar{t}H$ signal events passing the full top
reconstruction, along with backgrounds, is listed in
Table~\ref{tab:simplebb}. The accuracy of the reconstruction is
addressed in Appendix~\ref{app:recon}. With the strictest
set of cuts, giving the highest purity of top reconstruction, 50\% of
the tops we reconstruct in the signal are identifiable with a
parton--level top in the simulation.  However, at this stage of our
analysis, the reliable reconstruction of the top four momenta is not
yet the main goal. What we need are the un-associated $b$-jets, which
have to combine to the Higgs 4-momentum. Given the limited detector
resolution, we require the invariant mass of these two $b$-jets to lie
between 90 and 130~GeV. We find that, in the purest sample of
reconstructed tops, 68\% of the surviving signal events have the two
remaining $b$-jets in the ISR bucket that are correctly assigned (that
is, they correspond to parton--level $b$-quarks that originated in the
decay of the Higgs). The signal--to--background ratio for hard top
quarks can increase to around 1/10. Given the different
background uncertainties this number is promising. However, it
requires easily accessible side bands, in particular when we want to
extract the top Yukawa coupling from such a rate measurement.\bigskip

\begin{table}[b!]
\begin{tabular}{r|rrrr|r}
\hline
& $t\bar{t}H$ && $t\bar{t}b\bar{b}$ &$ bb\bar{b}\bar{b}jj$ & $S/B$ \cr
\hline
After acceptance cuts Eqs.\eqref{eq:global_cuts} and \eqref{eq:global_jets}
&1.197 & & 8.363 & 54.420 & 0.019 \cr
\hline
2 tops tagged, $\Delta\eta$ cuts & 0.587 & (0.125) & 2.762 & 10.654 & 0.044 \cr
$p_{T,t,1}>100$ GeV &0.485 & (0.111) & 2.392 & 8.364 & 0.045 \cr 
$p_{T,t,1}>200$ GeV &0.207 & (0.059) & 1.153 & 2.541 & 0.056 \cr 
$p_{T,t,1}>300$ GeV &0.064 & (0.021) & 0.405 & 0.507 & 0.071 \cr 
\hline
 \multicolumn{6}{c}{Mass window $m_{bb} = 90 - 130$~GeV} \cr \hline
2 tops tagged, $\Delta\eta$ cuts & 0.170 & (0.080) & 0.376 & 1.864 & 0.076 \cr 
$p_{T,t,1}>100$ GeV &0.144 & (0.072) & 0.317 & 1.396 & 0.084 \cr 
$p_{T,t,1}>200$ GeV &0.066 & (0.039) & 0.129 & 0.338 & 0.142 \cr 
$p_{T,t,1}>300$ GeV &0.021 & (0.015) & 0.034 & 0.043 & 0.276 \cr 
\hline
\end{tabular}
\caption{Cross sections (in fb) of events after successive selection
  cuts, as in Table \ref{tab:simplebb} but including the $\Delta\eta$ cuts
  of Eq.\eqref{eq:eta_cut}.}
\label{tab:simplebb_etacut}
\end{table}

The obvious choice of side bands is, of course, the invariant mass of
the two $b$-jets reconstructing the Higgs. While for the signal we
expect a peak around 125~GeV, possibly shifted towards lower values by
final state radiation escaping the momentum reconstruction, 
background (including combinatorics) can be well described by a 
log-normal distribution. In Figure~\ref{fig:mbb_left} we plot the
invariant mass of the two un-associated $b$-jets of events that have
survived the initial selection cuts of Eqs.\eqref{eq:global_cuts}
and~\eqref{eq:global_jets} and the top reconstruction algorithm. The
relatively narrow mass peak of the signal is well separated from a
broad feature of the backgrounds.  Not all signal events can be
associated with a parton--level Higgs momentum. In some cases the
reason is missing final state radiation, in others it can be due to the
$b$-jet combinatorics.

The situation significantly improves once we introduce a
$p_T^\text{min}$ cut on the reconstructed tops. Now the events under
the Higgs peak are more and more dominated by the actual Higgs signal, and
the signal peak separates cleanly from the broad maximum in the
background shape.  In Table~\ref{tab:simplebb} we show the rate of
signal and background events with $m_{bb}$ in the mass window of
90-130~GeV using the top reconstruction method to identify the Higgs
decay products. This can be directly compared to the naive
reconstruction methods from the previous section, summarized in
Table~\ref{tab:global}. Requiring that one of the two reconstructed top
quarks satisfy $p_{T,t}/m_t \gtrsim 1$ improves the
signal--to--background ratio by a factor of two.\bigskip

While at this stage the cut--and--count analysis is running out of
steam, it might be useful to show that the bucket reconstruction of
the two top decays gives us additional handles on the backgrounds. For
example, we can require the reconstructed momenta of the reconstructed
Higgs and tops to be central and not too widely separated in rapidity,
as is typically the case for heavy particle production,
\begin{equation}
\Delta \eta(t_1,t_2) < 3\; , \quad \Delta \eta(t_i,H) < 2 \;. 
\label{eq:eta_cut} 
\end{equation}
In Table~\ref{tab:simplebb_etacut} we show the corresponding signal
and background rates.  As can be seen, the ratio of signal to
background is greatly improved. Taking the most aggressive criteria,
requiring the leading reconstructed top to have $p_T > 300$~GeV and
the $\Delta \eta$ cut, we reach a $S/B = 1/3.6$, of which 70\% of the
signal $b$-jets in the mass window are correctly identified from the
Higgs decay. In general, this shows that the top reconstruction
provides two handles that can improve the signal strength. First, it
gives us a more accurate method to assign $b$-jets to the Higgs decay,
reducing the combinatorial background. Second, it gives us kinematic
information for the event that can be exploited to discriminate signal
events.\bigskip

Clearly, the proposed bucket analysis is unlikely to be the last
experimental word in extracting purely hadronic $t\bar{t}H$ events
from QCD background. However, our analysis shows that QCD and
combinatorial backgrounds do not render this channel
hopeless. Reconstructing the top decay products preferably in the
slightly boosted regime can solve both problems and even leave the
analysis with simple side bands, like the $m_{bb}$ distribution.

\section{Conclusion}
\label{sec:conclusion}

We have demonstrated a method to extract the associated production of
the Higgs along with top pairs in the {\sl fully hadronic channel},
using the top buckets method or Ref.~\cite{buckets} to reconstruct the
hadronic tops. Using this reconstruction technique nets us several
useful advantages over more naive methods to reduce the very large
backgrounds.

First, by having an accurate method of determining which two of the
four $b$-jets in the event should be assigned to the top pair, we can
cut through the combinatorial problem of identifying the two $b$-jets
from the Higgs decay.  Side bands with $m_{bb} \lesssim 100$~GeV or
$m_{bb} \gtrsim 200$~GeV can be used to determine the background shape
and extract the background cross section after cuts.  Second, the
bucket algorithm not only identifies the $b$-jets from the top decay,
it also gives a good approximation of the top and Higgs momenta.  This
allows us to place additional cuts, for example on the transverse
momenta of the top quarks or on the $\Delta \eta$ of the various
parton--level objects. Both of them help to reject background. In
particular, requiring a small boost of the top quarks eases the
combinatorial problem~\cite{tth_bb}.  Further, more detailed analyses
may improve on the fairly crude cuts we have chosen in this
proof--of--concept paper.\bigskip

In this paper, we concentrated on demonstrating the stability of our
reconstruction technique despite the potentially large simulation and
theoretical uncertainties inherent to a QCD background consisting of
four $b$-jets with many extra un-tagged jets. Using both
\textsc{Alpgen} and \textsc{Sherpa}, we have validated that the
simulation issues are under control. However, our study also clearly
shows that the theory uncertainties on this kind of backgrounds are
hardly covered by a factor two on the rate prediction.  These issues
can be mitigated in the experiments by use of the ample side-bands
that this analysis affords. In addition to the background dominated
regions outside of the Higgs mass window, there are also many
sidebands available, for example in the distribution of jet
multiplicity.\bigskip

\begin{center}
{\bf Acknowledgments}
\end{center}

MB and TP would like to thank the Aspen Center of Physics because the
idea for this paper was born on a Snowmass ski lift. TP would
furthermore like to thank the CCPP at New York University for their
hospitality, which added Washington Square as a second location
crucial to the progress of this paper. Fermilab is operated by Fermi
Research Alliance, LLC, under contract DE-AC02-07CH11359 with the
United States Department of Energy.

\newpage

\appendix

\section{Signal and background simulations}
\label{app:QCD}

In this Appendix we will confirm that the analysis described in this
paper does not critically depend on uncertainties in the way we
compute our signal and backgrounds. For the signal and the
$t\bar{t}b\bar{b}$ background we primarily rely on
\textsc{Sherpa}~\cite{sherpa} predictions with up to one
additional hard jet merged using the \textsc{Ckkw}
approach~\cite{ckkw}. For the $t\bar{t}H$ signal we test our results
using \textsc{Madgraph}~\cite{madgraph}, with up to one hard jet
included in the \textsc{Mlm} scheme~\cite{mlm}. Both event samples are
normalized to the next-to-leading order rate (extrapolated to 13 TeV) of
504~fb~\cite{tth_nlo}, times a Higgs branching ratio of 57.7\%.
This corresponds to 129~fb for the purely hadronic decay channel. In
Table~\ref{tab:app_signal} we observe a small difference in the
normalization of the two event samples. The reason is that as a cross
check in the \textsc{Madgraph} simulation, we do not require hadronic
top decays in the simulation. As a result, the decays to hadronic taus
contribute to the signal. Our default \textsc{Sherpa} simulation conservatively does
not include these events.\bigskip

For the $t\bar{t}b\bar{b}$ background we test the \textsc{Sherpa}
simulation with up to one hard additional jet with an
\textsc{Alpgen}~\cite{alpgen} simulation without additional hard
jets. Again, both samples are normalized to the next-to-leading order
rate of 1037~fb~\cite{ttbb_nlo} after the generator cuts $p_{T,b} >
35$~GeV, $|\eta_b|<2.5$, and $\Delta R_{bb}>0.9$.  This rate is
approximate because, in the absence of a next-to-leading order
prediction for $\sqrt{s} = 13$~TeV, we are forced to first extract the
$K$ factor for 14~TeV and the cuts of Ref.~\cite{ttbb_nlo}, including
a regularizing cut on the invariant mass of the two bottom
quarks. We then multiply our cross section at 13~TeV by this $K$
factor. This approach is not ideal, but better then just using the
leading order prediction.

In Table~\ref{tab:app_signal} we see that the transverse momentum
distributions for the ``reconstructed'' top quarks (which, for $bb\bar{b}\bar{b}$, 
do not correspond to any parton-level tops) from \textsc{Alpgen} are softer
than for \textsc{Sherpa}. This effect comes from the generically harder
jets of \textsc{Ckkw} merging, compared to those from the parton shower.  In
order to be conservative, we use the merged \textsc{Sherpa} results
for our analysis. On the other hand, the difference of less than 20\%
is well within the theory uncertainties for this background.\bigskip
 
As argued in Section~\ref{sec:multijets}, the most dangerous background
events should be correctly described by our \textsc{Alpgen} simulation
of the $bb\bar{b}\bar{b}jj$ background plus \textsc{Pythia} parton
shower, as the required extra jets must be hard, and therefore well-modeled by 
the matrix-level process. With our merged \textsc{Sherpa} simulation of
$bb\bar{b}\bar{b}+0/1$~jet, we test several aspects of our main background
simulation:
\begin{enumerate} 
\item We check if the events with two additional hard jets are
  indeed the leading background after the kind of global cuts
  proposed in Section~\ref{sec:multijets}. This aspect is very
  important for the appropriate simulation of the QCD background in an
  actual analysis.
\item We test if our analysis depends on the simulation of the second
  un-tagged jet either with the hard matrix element or through the
  parton shower. In this way we can estimate an important source of
  theory uncertainties.
\item As a measure of the level of agreement between the two
  simulations we compute the merged \textsc{Sherpa} event rate with a
  consistent variation of the renormalization and factorization
  scales. Ideally, the two simulations should agree
  within this scale variation in the signal region of the buckets
  analysis. Because the merged \textsc{Sherpa} prediction includes
  some leading next-to-leading order contribution such a numerical
  agreement also indicates that our $bb\bar{b}\bar{b}jj$ simulations
  should not be plagued by huge QCD corrections.
\end{enumerate} 
This extensive list of tests should give a clear answer to the
question if purely hadronic $t\bar{t}H$ searches can be done in the
presence of the large QCD backgrounds. Finally, we point out that
merged \textsc{Sherpa} simulations can define excellent side bands in
the $n_j$ distribution~\cite{jet_scaling}, which together with side
bands in $m_{bb}$ should be sufficient to control the background rate
in the signal region in an experimental analysis. Our results suggest
that such an approach would require a merged simulation of
$bb\bar{b}\bar{b}$ with up to at least two hard light-flavor or gluon jets,
which is beyond our CPU capabilities.\bigskip

\begin{table}[t]
\begin{tabular}{l|rr|rr}
\hline
& \multicolumn{2}{c}{$t\bar{t}H$}
& \multicolumn{2}{c}{$t\bar{t}b\bar{b}$} \cr \hline
& \textsc{Madgraph} (merged) & \textsc{Sherpa} (merged)
& \textsc{Alpgen} (shower) & \textsc{Sherpa} (merged) \cr \hline
After acceptance Eqs.\eqref{eq:global_cuts} and \eqref{eq:global_jets}
&1.390 &1.197  & 7.903 &8.363  \cr
\hline
 2 tops tagged & 1.100 &0.894& 5.893 & 5.882   \cr
$p_{T,t,1} > 100$~GeV &0.866 & 0.709 & 4.684 & 4.868  \cr
$p_{T,t,1} > 200$~GeV & 0.342 & 0.289 & 1.806 & 2.189  \cr
$p_{T,t,1} > 250$~GeV & 0.180 & 0.165 & 0.978 & 1.295  \cr
$p_{T,t,1} > 300$~GeV & 0.092 & 0.089 & 0.502 & 0.724  \cr
\hline
& \multicolumn{4}{c}{Mass window $m_{bb} = 90 - 130$~GeV} \cr \hline
 2 tops tagged & 0.337 &  0.259 & 1.016 & 0.859  \cr
$p_{T,t,1} > 100$~GeV & 0.274 &0.208 & 0.780 & 0.688 \cr
$p_{T,t,1} > 200$~GeV & 0.112 & 0.091 & 0.260 & 0.265  \cr
$p_{T,t,1} > 250$~GeV & 0.058 & 0.050 & 0.128 & 0.144 \cr
$p_{T,t,1} > 300$~GeV & 0.032 & 0.028 & 0.059 & 0.072 \cr
\hline
\end{tabular}
\caption{Signal and background cross sections (in fb) after successive
  selection cuts, showing the different ways of simulating the
  signal and the irreducible $t\bar{t}b\bar{b}$ background. All
  conventions correspond to the final result shown in
  Table~\ref{tab:simplebb}. We use the \textsc{Sherpa} results for our
  main analysis.}
\label{tab:app_signal}
\end{table}

\begin{figure}[b!]
\includegraphics[width=0.32\textwidth]{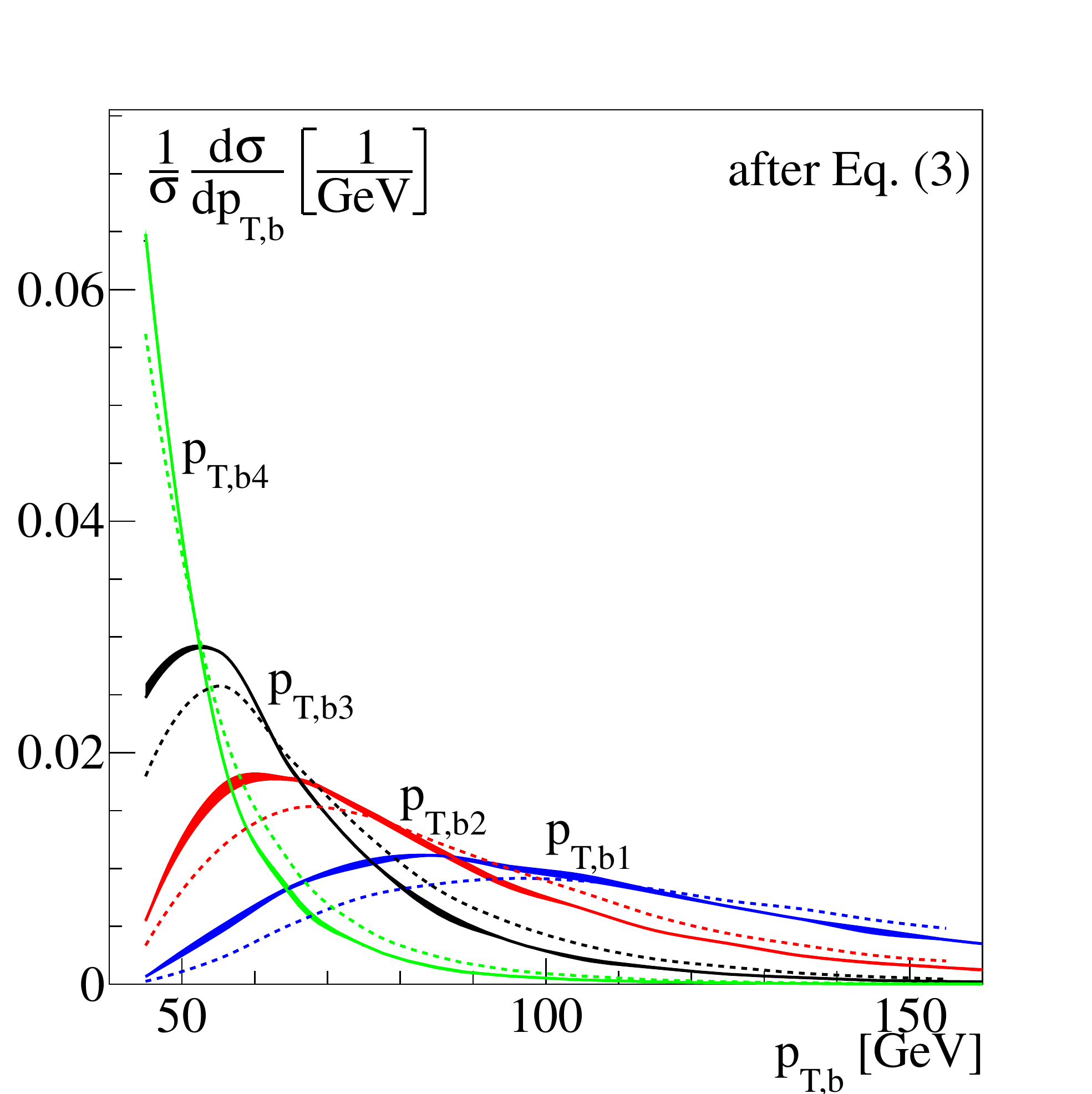}
\hspace*{0.1\textwidth}
\includegraphics[width=0.32\textwidth]{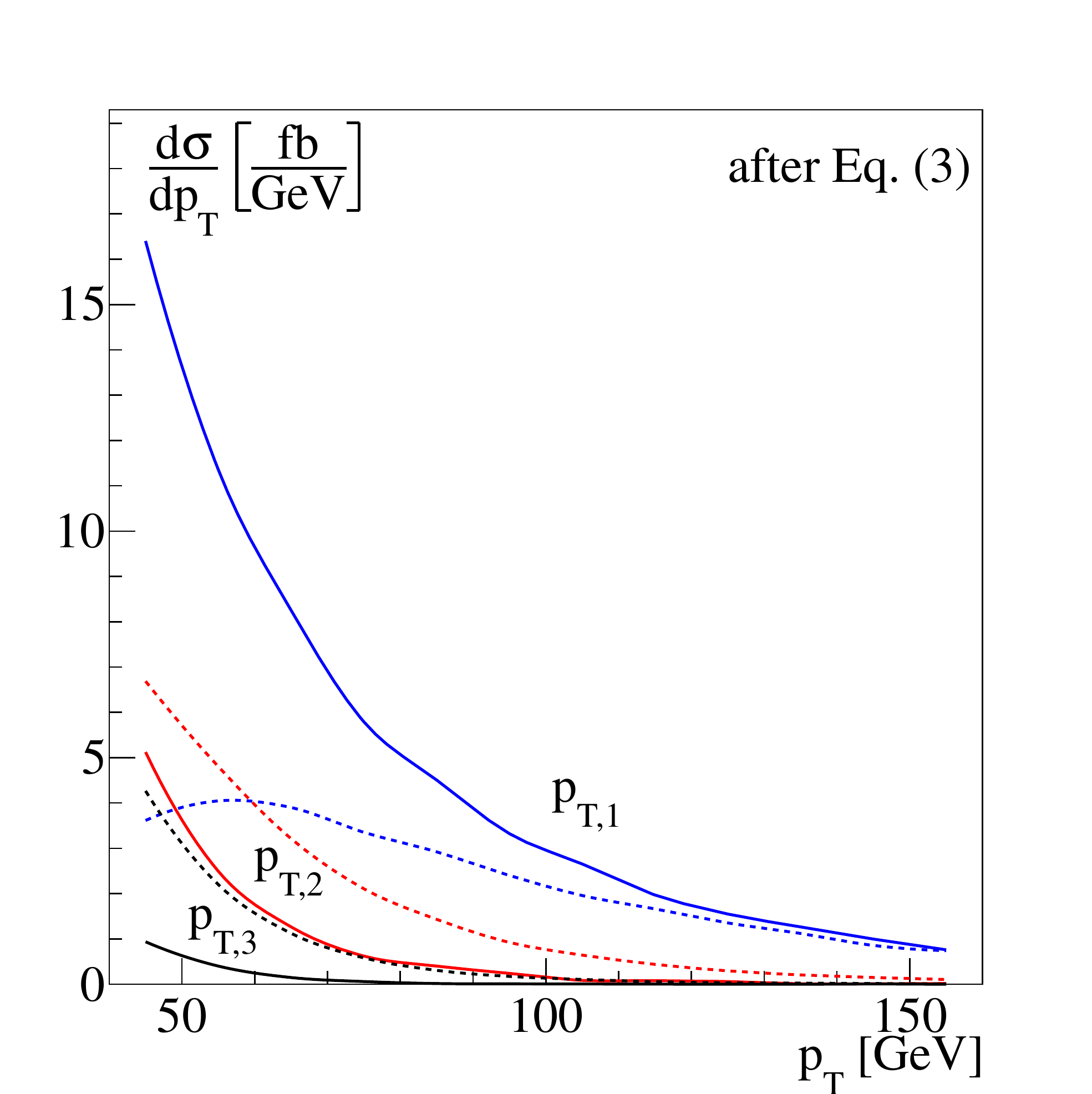}
\caption{Normalized transverse momentum distributions of four
  $b$-tagged jets (left) and the leading three additional un-tagged
  jets (right). All events include four hard $b$-jets according to
  Eq.\eqref{eq:global_cuts}, but no requirement on the number of additional
  un-tagged jets.  The solid curves correspond to the merged
  $bb\bar{b}\bar{b}+0/1$~jets simulation with \textsc{Sherpa} while
  the dashed curves show the $bb\bar{b}\bar{b}jj$ events from
  \textsc{Alpgen}. The scale variation for the \textsc{Sherpa} result
  is indicated by the widths of the solid lines in the left panel.}
\label{fig:app_pt}
\end{figure}

In Figure~\ref{fig:app_pt} we first show the normalized transverse
momenta of the four $b$-jets. The curves are set to unit
normalization, as the significantly different cross section of the
$bb\bar{b}\bar{b}jj$ and the merged $bb\bar{b}\bar{b}+0/1$~jets is
almost entirely due to the different number of un-tagged jets in the
events. In the left panel we see that the leading $b$-jet agrees in
the two approaches, while the second to fourth $b$-jets become
increasingly harder in the \textsc{Alpgen} $bb\bar{b}\bar{b}jj$
sample. This is because, with two additional hard jets, the available
recoil momentum is slightly larger. The sensitivity to the proper
simulation of the recoil is also the reason why the \textsc{Alpgen}
curves are not covered by the scale variation of the
\textsc{Sherpa} simulation.

The results for the leading un-tagged jets in the right panel of
Figure~\ref{fig:app_pt} look much less promising. The very different
integrated rates under the curves reflect the additional events with
only $bb\bar{b}\bar{b}$ in the hard process plus any number of parton
shower jets. This is particularly obvious for the first un-tagged jet,
where the \textsc{Sherpa} simulation includes a majority of events
with only one additional jet while the \textsc{Alpgen} sample will
always include a second hard jet together with the first. For the
second un-tagged jet the integrated rates in the $p_{T,j}$
distributions are similar for the two samples. The \textsc{Alpgen}
simulation gives a significantly harder second jet from the matrix
element while the second jet in our \textsc{Sherpa} sample
(corresponding to the first jet from the parton shower) tends to be
soft. The third un-tagged jet is the second parton shower jet in our
\textsc{Sherpa} sample, while in the \textsc{Alpgen} simulation it is
the first parton shower jet radiated from a harder core process. Both
effects combined result in a significantly harder $p_{T,j}$ spectrum
for the \textsc{Alpgen} sample. These distribution suggest that if our
signal region should indeed require two or even three hard un-tagged
jets to mimic top decay jets, the $bb\bar{b}\bar{b}jj$ sample from
\textsc{Alpgen} should be the appropriate, conservative
estimate.\bigskip

\begin{figure}[t]
\includegraphics[width=0.32\textwidth]{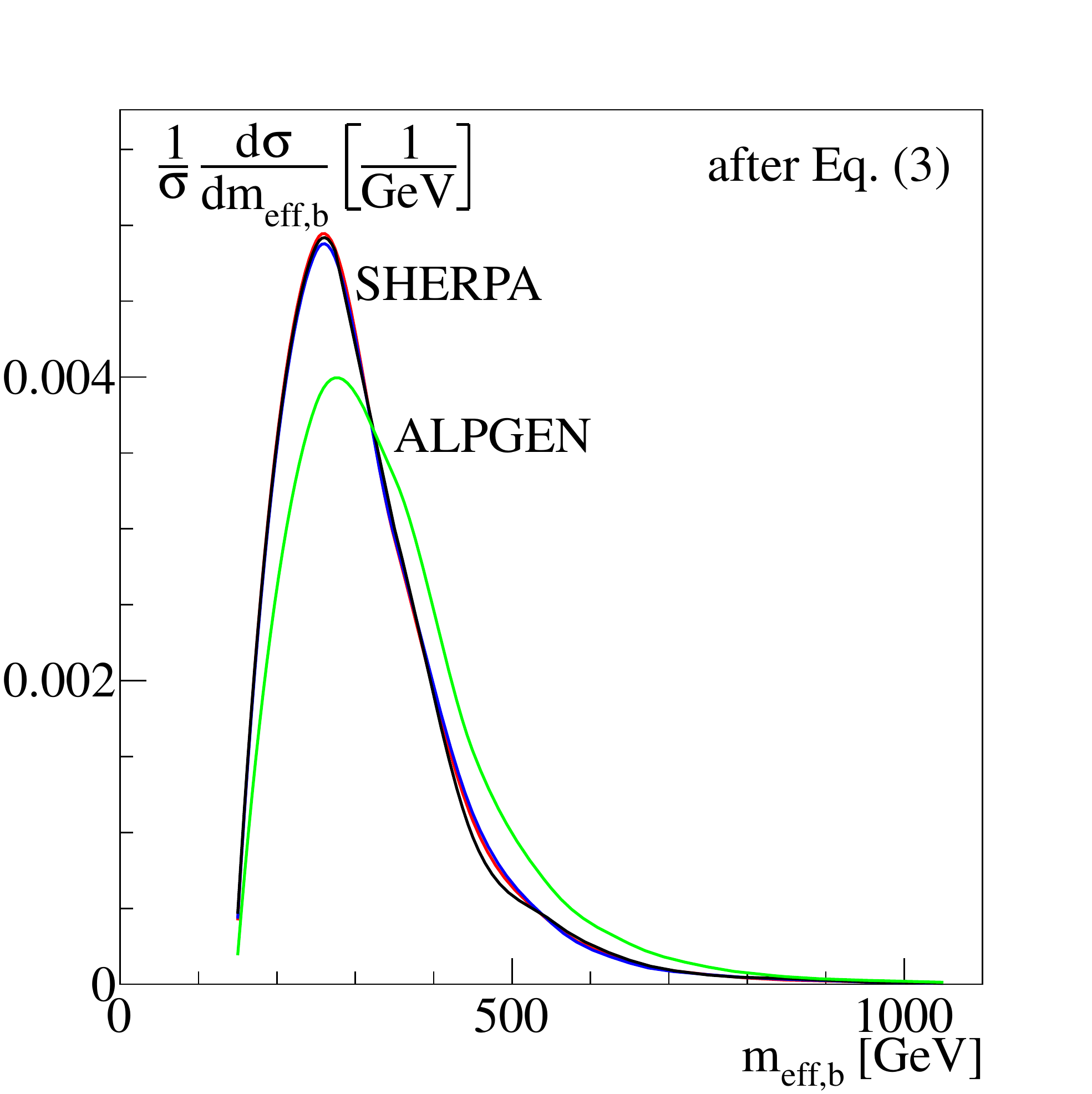}
\includegraphics[width=0.32\textwidth]{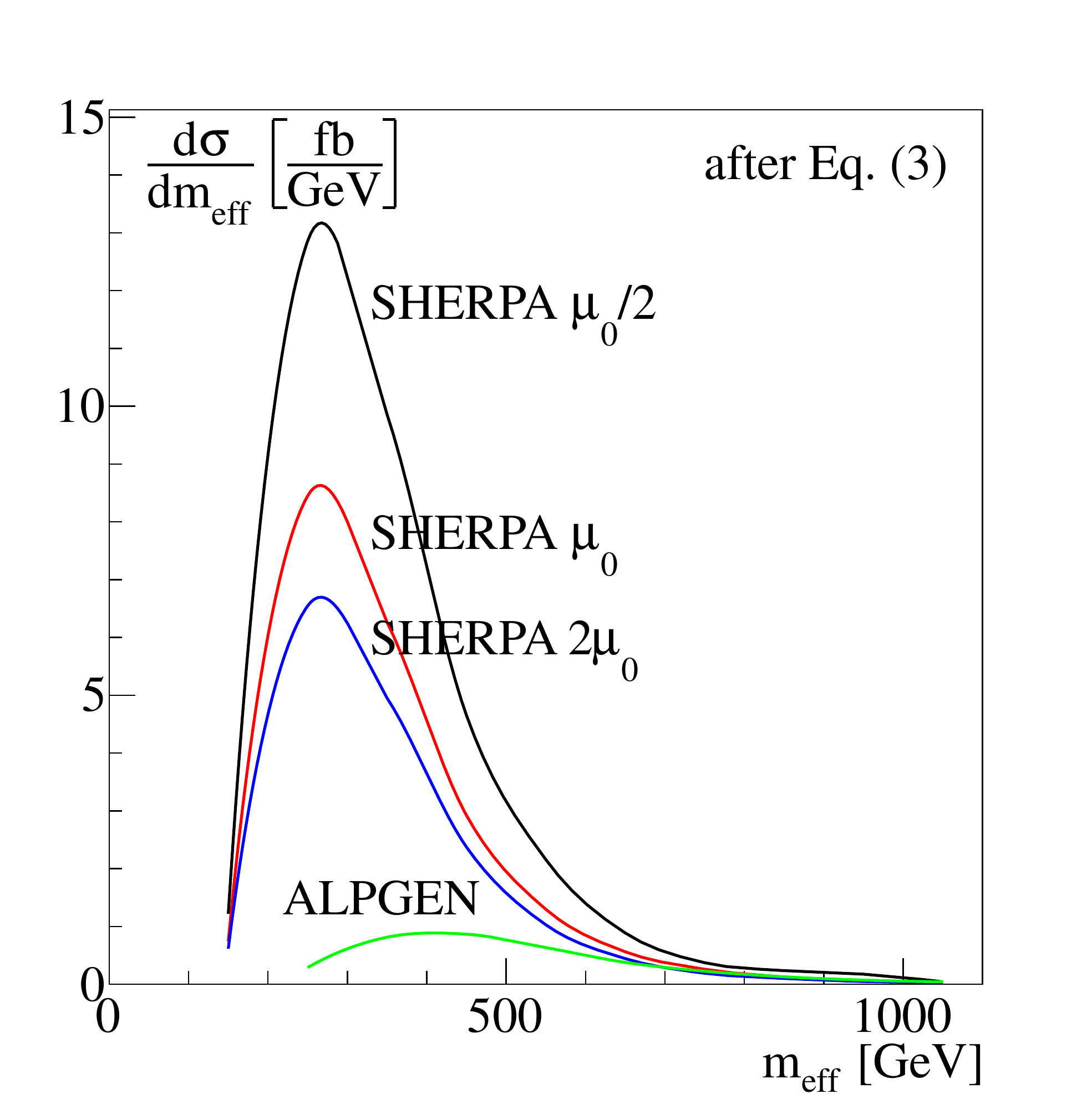}
\includegraphics[width=0.32\textwidth]{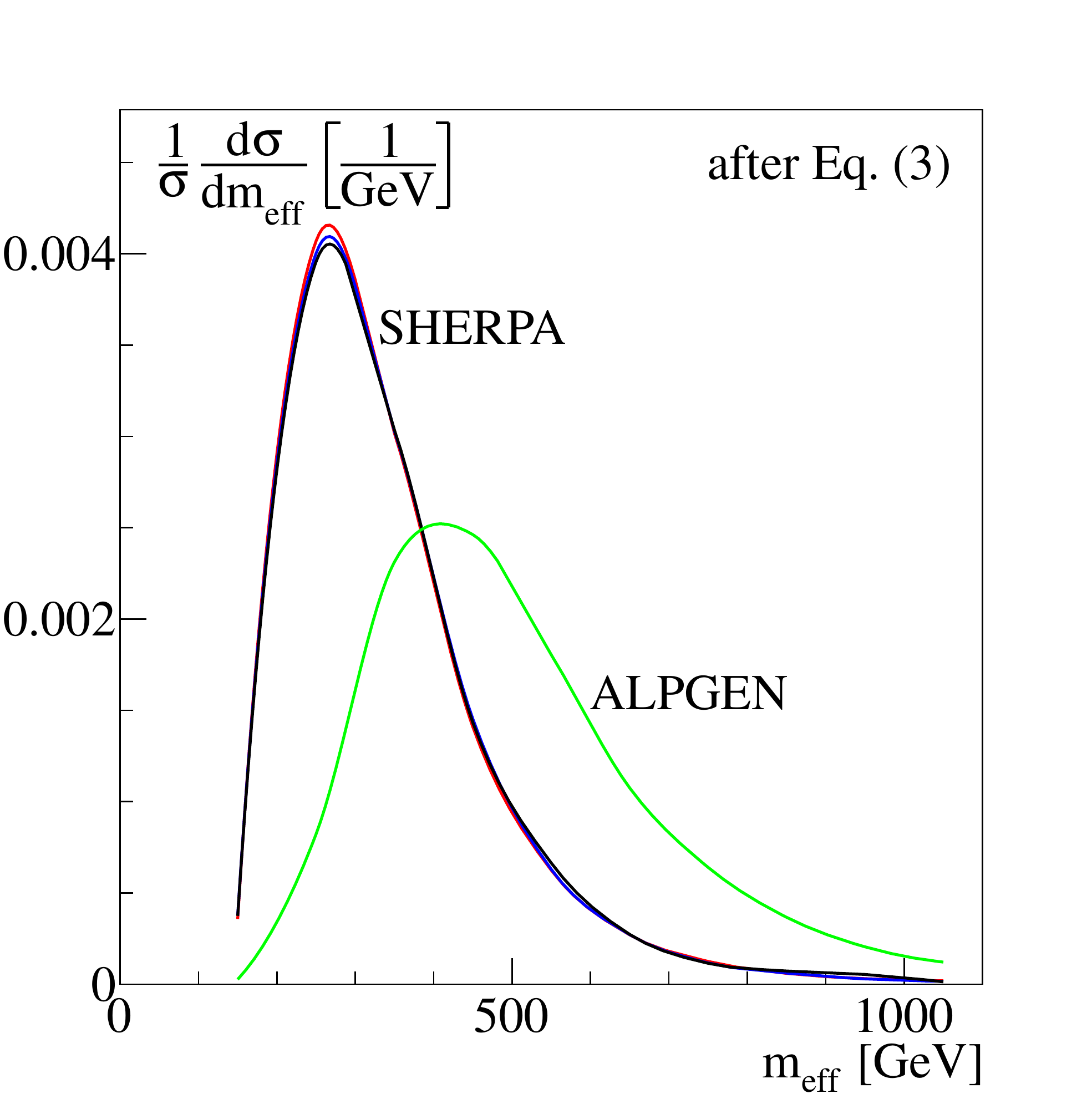} \\
\includegraphics[width=0.32\textwidth]{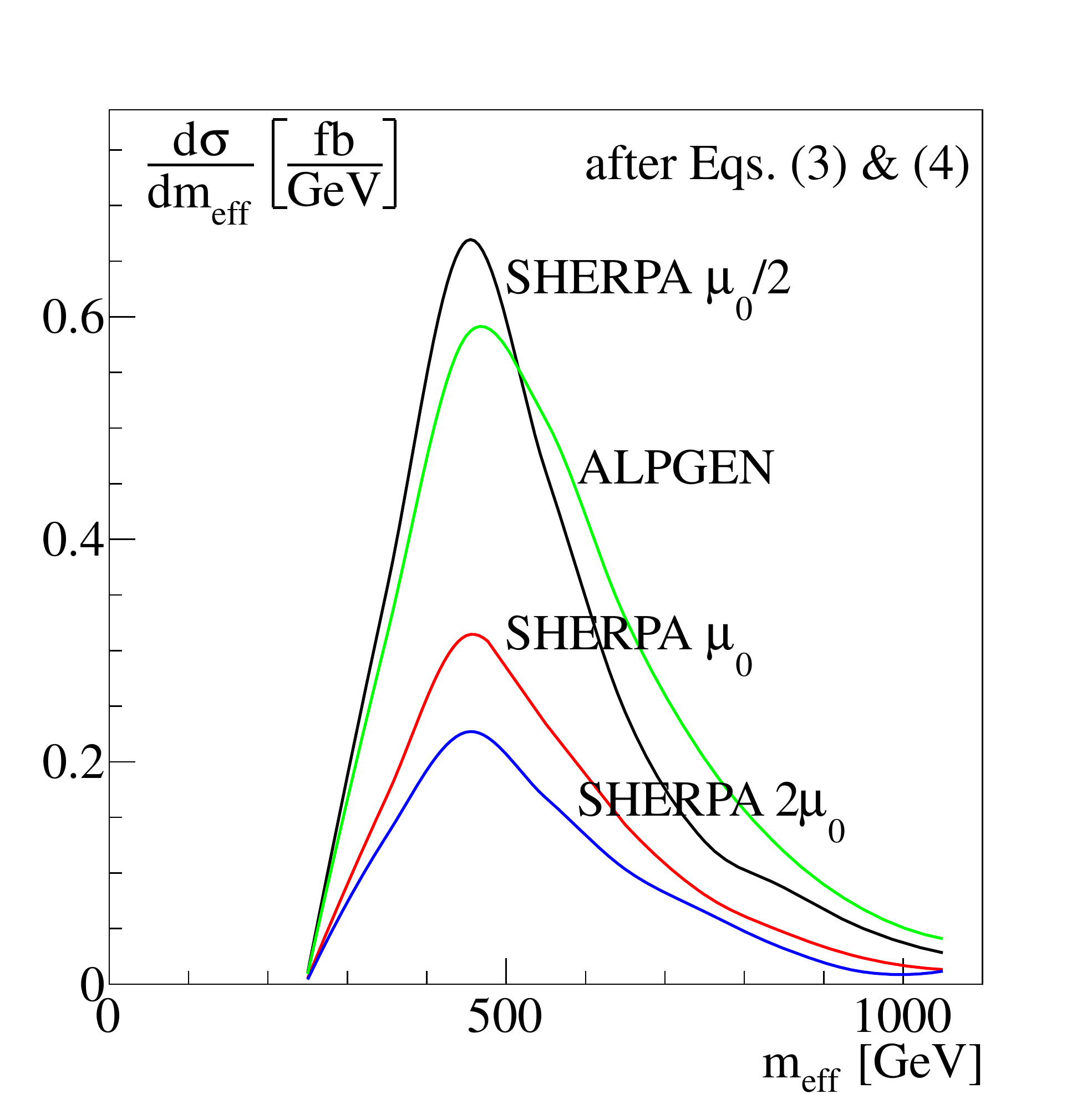}
\includegraphics[width=0.32\textwidth]{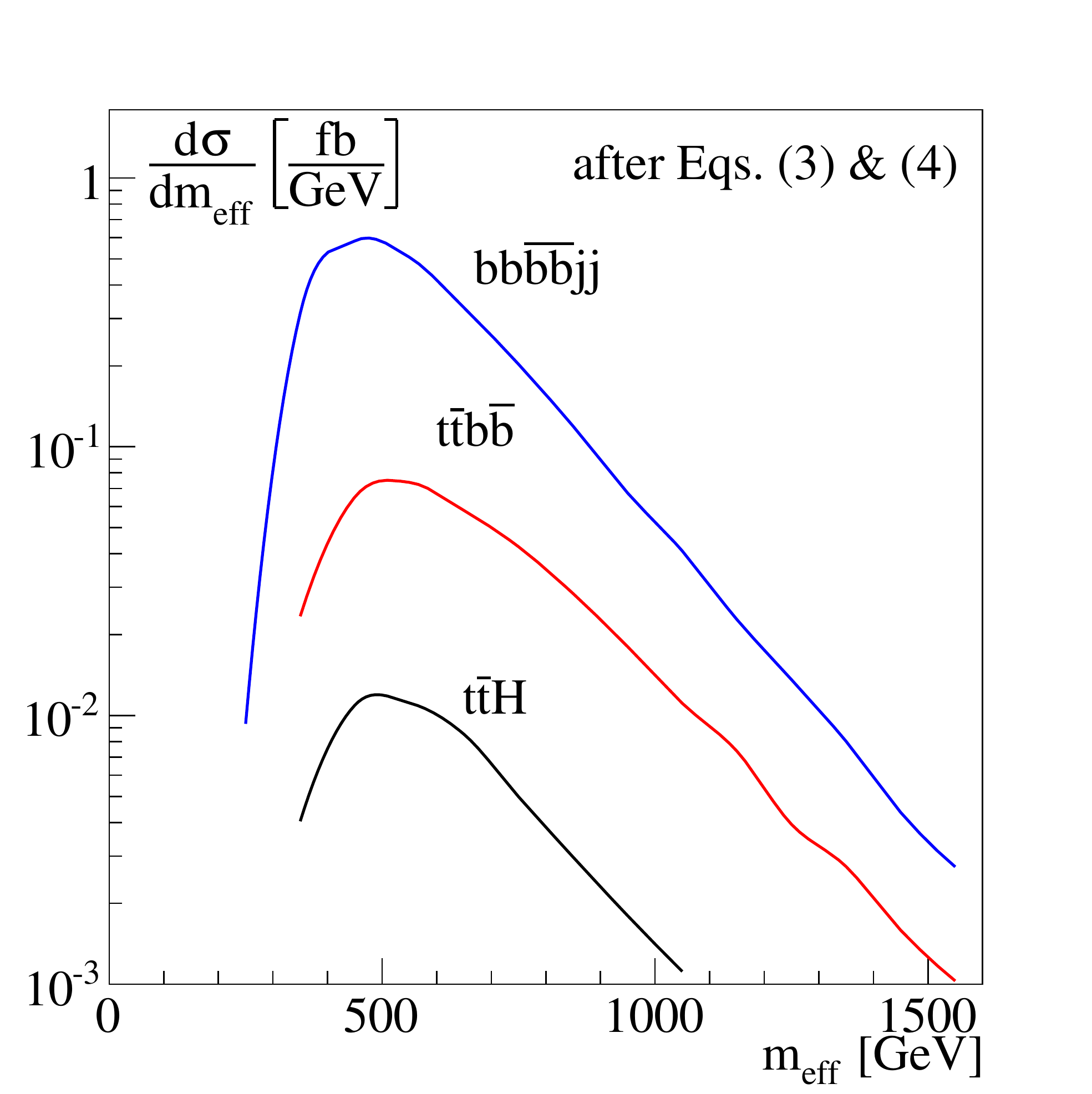}
\includegraphics[width=0.32\textwidth]{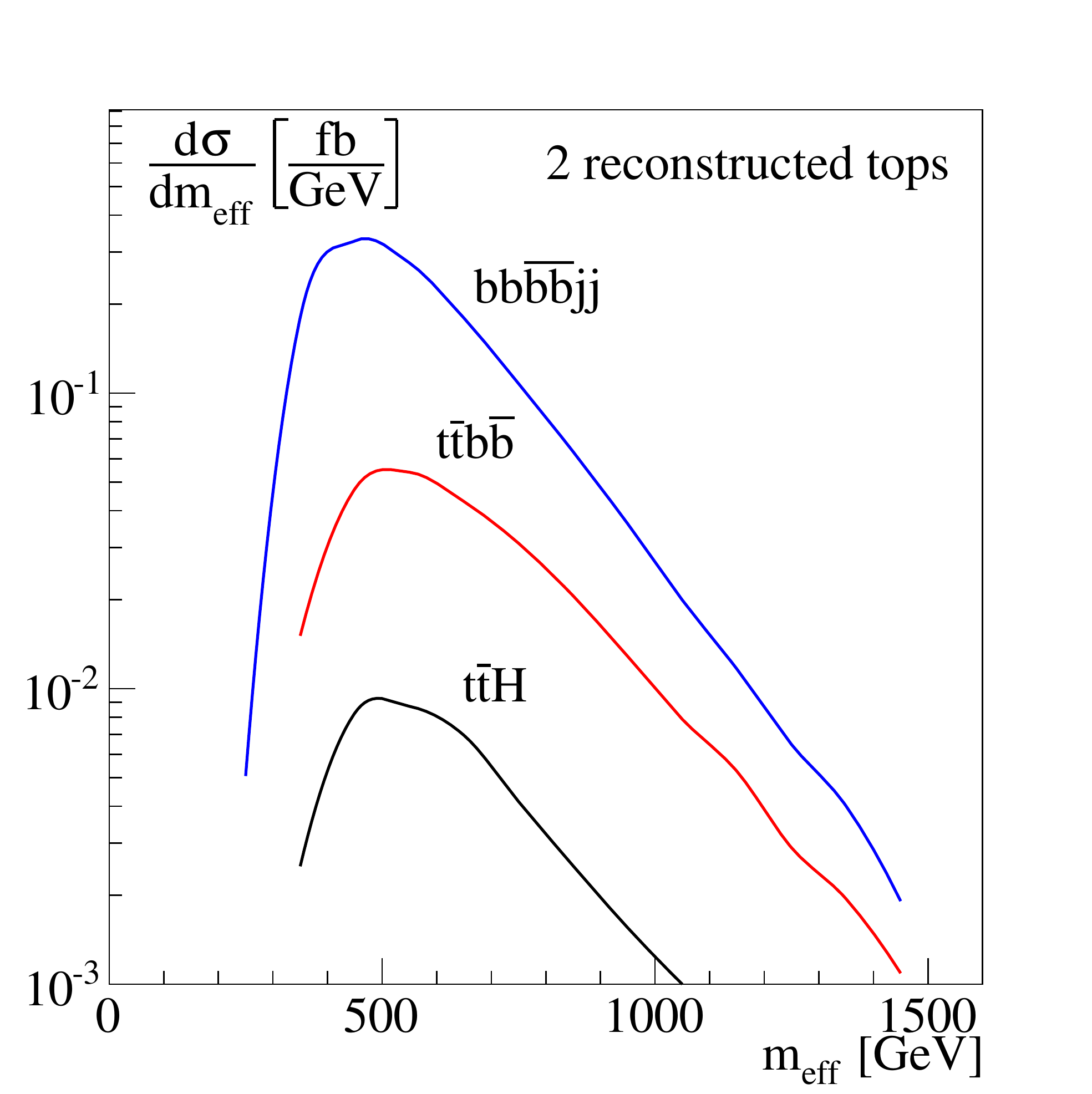}
\caption{Effective mass distributions. In the upper panels we require
  four hard $b$-jets according to Eq.\eqref{eq:global_cuts}, but no
  cut on the number of additional un-tagged jets.  We show
  the normalized $m_{\text{eff},b}$ distribution for four $b$-jets (left), as
  well as the all-jet $m_\text{eff}$ from \textsc{Sherpa} and
  \textsc{Alpgen} normalized to the rates (center) and normalized to
  unity (right). In the lower panels we require two additional
  un-tagged jets fulfilling Eq.\eqref{eq:global_jets}. Here we show
  the \textsc{Alpgen} vs \textsc{Sherpa} results (left), our default
  signal and background samples (center), and the default signal and
  backgrounds for all events with two valid top buckets (right).}
\label{fig:app_meff} 
\end{figure}

In the upper panels of Figure~\ref{fig:app_meff} we show the two
relevant effective mass variables defined in
Eq.\eqref{eq:def_meff}. We require four $b$-jets according to
Eq.\eqref{eq:global_cuts} and any number of un-tagged jets after
Eq.\eqref{eq:global_jets}.  Unlike Figure~\ref{fig:global} we now only
show the different results for the $bb\bar{b}\bar{b}$ background. In
the upper left panel we again show the normalized observable from the
multi-$b$ sector. The conclusion follows from the discussion of the
transverse momenta of the four $b$-jets: the dependence of the
$m_{\text{eff},b}$ distribution on the simulation is small, clearly
when we look at the \textsc{Sherpa} scale variation, but also in terms
of the difference between \textsc{Alpgen} and \textsc{Sherpa}.  The
only difference is that the $bb\bar{b}\bar{b}jj$ simulation with
\textsc{Alpgen} predicts slightly harder multi-$b$ sectors.

In the upper central panel of Figure~\ref{fig:app_meff} we see that in the
background region the difference in rate between the two simulations
is again dramatic, and certainly not covered by the scale variation of
the $bb\bar{b}\bar{b}+0/1$~jets simulation with \textsc{Sherpa}. On
the other hand, this result is entirely expected, and the difference
becomes increasingly smaller once our analysis of the signal region
requires something like $m_\text{eff} > 500$~GeV.  Increasing the cut
to $m_\text{eff} > 700$~GeV brings them into agreement within scale
uncertainties.  In that regime the bulk of the
$b\bar{b}b\bar{b}+0/1$~jet events do not contribute, so the two
simulations should roughly agree within the scale variation of the
\textsc{Sherpa} simulation.  The upper right panel confirms that in
 the \textsc{Alpgen} simulation with the
$bb\bar{b}\bar{b}jj$ hard process also gives a harder spectrum in $m_\text{eff}$. 

In the lower panels of Figure~\ref{fig:app_meff} we not only require
four $b$-jets following Eq.\eqref{eq:global_cuts}, but also at least
two un-tagged jets fulfilling Eq.\eqref{eq:global_jets}. Two such
additional jets are implicitly required for any event passing the
bucket analysis described in Section~\ref{sec:buckets}. First, we see
in the left panel that simply asking for two un-tagged jets
suppresses the central prediction from the merged
$bb\bar{b}\bar{b}+0/1$~jets simulation to roughly half the
$bb\bar{b}\bar{b}jj$ prediction. Both $m_\text{eff}$ distributions
peak around 500~GeV, and the \textsc{Alpgen} rate is covered by the
scale variation of the \textsc{Sherpa} simulation. In the central
lower panel we compare the distributions for our default signal and
background simulations after requiring four $b$-tagged and two
un-tagged jets. Finally, in the lower right panel we show the same
distribution for all events passing the bucket analysis described in
Section~\ref{sec:buckets}. As compared to the acceptance cuts,
Eqs.\eqref{eq:global_cuts} and~\eqref{eq:global_jets}, there is hardly
any change, which means that the improvements by the bucket analysis
 are more promising than the global cuts proposed in
Section~\ref{sec:multijets}, with the added advantage of avoiding
shaping the background distributions, such as $m_\text{eff}$.\bigskip

\begin{table}[t]
\begin{tabular}{l|r|r|rrr}
\hline
& $t\bar{t}H$ \textsc{Sherpa} 
& $bb\bar{b}\bar{b}jj$ \textsc{Alpgen} 
& \multicolumn{3}{c}{$bb\bar{b}\bar{b}$+jets \textsc{Sherpa}} \cr
&&& $2\mu_0$ & $\mu_0$ & $\mu_0/2$ \cr
\hline
After acceptance Eqs.\eqref{eq:global_cuts} and \eqref{eq:global_jets}
&1.197  & 54.420 & 18.825 & 25.812 & 50.974\cr
\hline
2 tops tagged  & 0.894 & 29.356 &7.507 & 10.091 & 20.473 \cr
$p_{T,t,1} > 100$~GeV & 0.709 &  20.838 & 5.049  & 7.283 &13.843  \cr
$p_{T,t,1} > 200$~GeV & 0.289 &  5.194 &1.155 & 1.419 &3.018  \cr 
$p_{T,t,1} > 250$~GeV & 0.165 & 2.213  &0.488  & 0.717 &1.361\cr 
$p_{T,t,1} > 300$~GeV & 0.089 &  0.917&0.218  & 0.351 & 0.645 \cr
\hline
& \multicolumn{5}{c}{Mass window $m_{bb} = 90 - 130$~GeV} \cr \hline
2 tops tagged & 0.259 & 5.424& 1.143 & 1.726 &3.354 \cr
$p_{T,t,1} > 100$~GeV & 0.208 &  3.600 &0.749 & 1.111 & 2.118 \cr
$p_{T,t,1} > 200$~GeV & 0.091 & 0.679 &0.133  & 0.161 & 0.233 \cr
$p_{T,t,1} > 250$~GeV & 0.050 & 0.233 &0.031  & 0.044 &0.082 \cr
$p_{T,t,1} > 300$~GeV & 0.028 & 0.082 &0.020  & 0.015 & 0.041 \cr
\hline
\end{tabular}
\caption{Cross section (in fb) for signal and background events after successive
  selection cuts, showing the different ways of simulating the
  $bb\bar{b}\bar{b}$ background. All conventions correspond to the
  final result shown in Table~\ref{tab:simplebb}. Reconstructed tops for $bb\bar{b}\bar{b}$
  do not correspond to any real parton-level object.}
\label{tab:app_bbbb}
\end{table}

From the above comparison we expect that for an actual top and Higgs
analysis the two background simulations should be fairly consistent
once we probe sufficiently hard multi-jet configurations. The scale
uncertainty of our \textsc{Sherpa} simulation determines the numerical
level of this consistency.  Moreover, in the signal phase space region
the $bb\bar{b}\bar{b}jj$ background simulation with parton shower
should predict larger backgrounds and give us a conservative estimate.
In Table~\ref{tab:app_bbbb} we show the different $bb\bar{b}\bar{b}$
background rates after the buckets analysis. Indeed, the simulations
agree roughly within the sizable scale uncertainties.  The
$b\bar{b}b\bar{b}jj$ simulation with \textsc{Alpgen} gives the largest
rate, in particular once we require large, signal-like $m_{bb}$ values
and sizable $p_T$ of the fake reconstructed top buckets. For the experimental
analysis this implies firstly that the signal-to-background ratio for
purely hadronic $t\bar{t}H$ events at 13~TeV can be of the order
1/3. Second, the remaining number of signal events will be an issue
for a cut-and-count analysis. Lastly, the uncertainties on the
background simulation will require a careful background determination
from side bands and control regions. Any kind of $t\bar{t}H$ analysis
which does not provide at least a slight mass peak in the $m_{bb}$
distribution around 125 GeV would have a hard time convincing the
authors of this study. Our buckets analysis will carefully ensure that
this simple side band is clearly visible, in spite of the fact that
this requirement might lead to a slightly reduced performance of our
analysis.

\section{Top reconstruction}
\label{app:recon}

In this Appendix, we provide metrics concerning the reconstruction of
the top pair, using the bucket method originally proposed in
Ref.~\cite{buckets}, and described in detail in
Section~\ref{sec:buckets}. As every event contains exactly four
$b$-jets, reconstructing the tops leaves two $b$-jets that can be
identified as coming from the decay of the Higgs, thus we can
reconstruct its momentum as well. While cuts on the top and Higgs
kinematics are not critical to the analysis presented in this paper,
for example a multivariate version of the same analysis would
immediately be able to benefit from a valid bucket
reconstruction.\bigskip

\begin{figure}[t]
\includegraphics[width=0.30\textwidth]{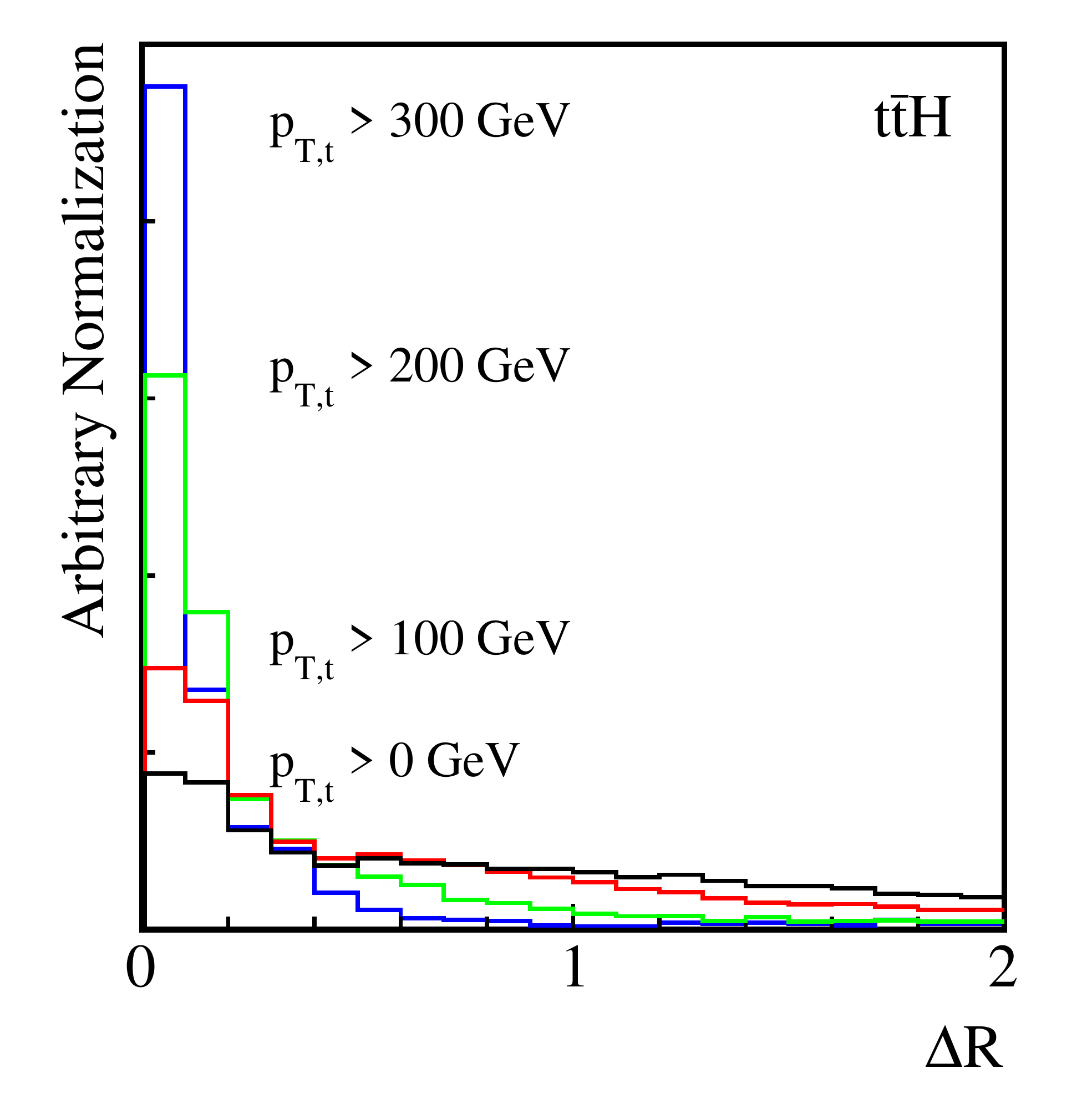}
\hspace*{0.02\textwidth}
\includegraphics[width=0.30\textwidth]{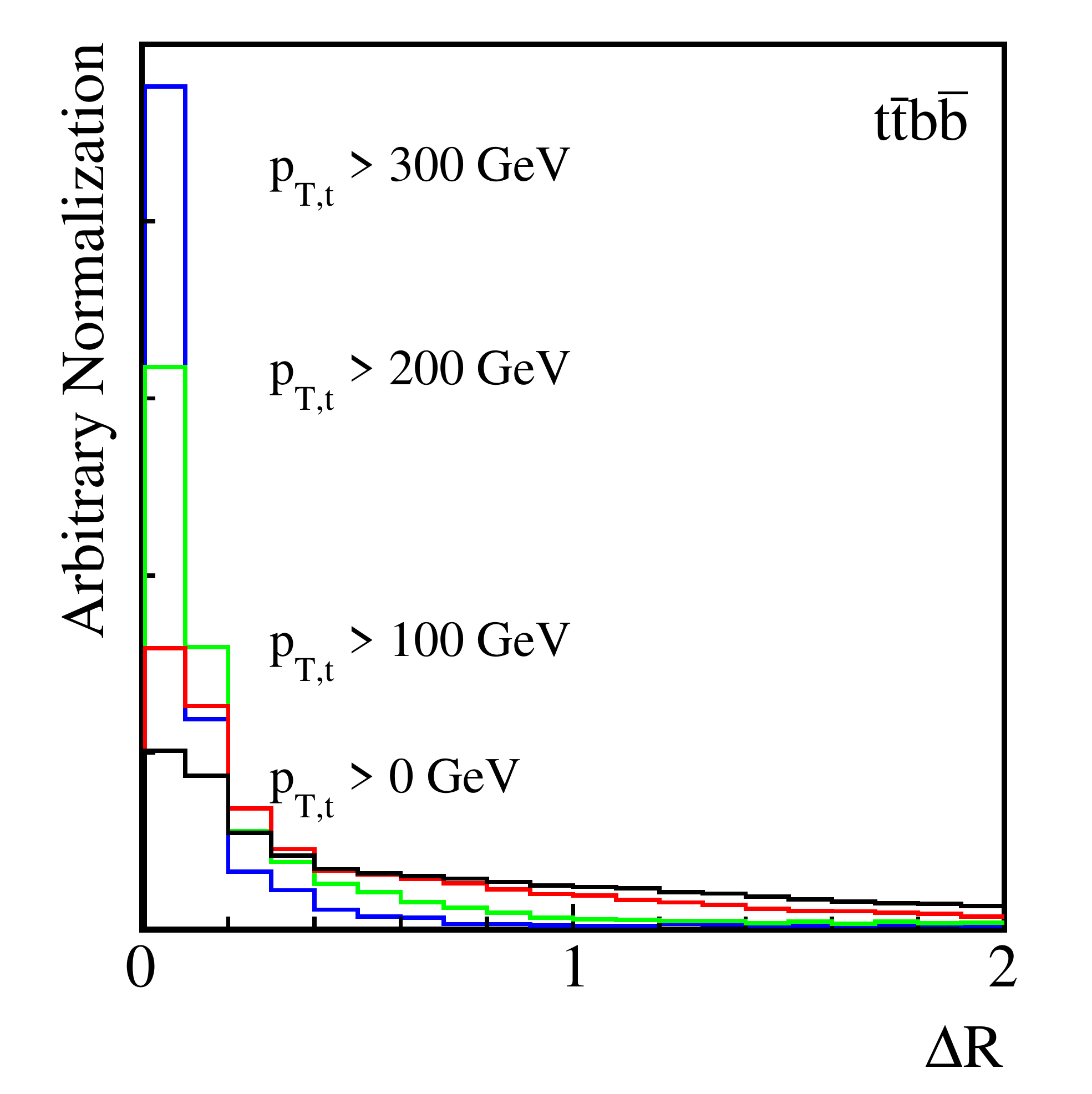}
\hspace*{0.02\textwidth}
\includegraphics[width=0.30\textwidth]{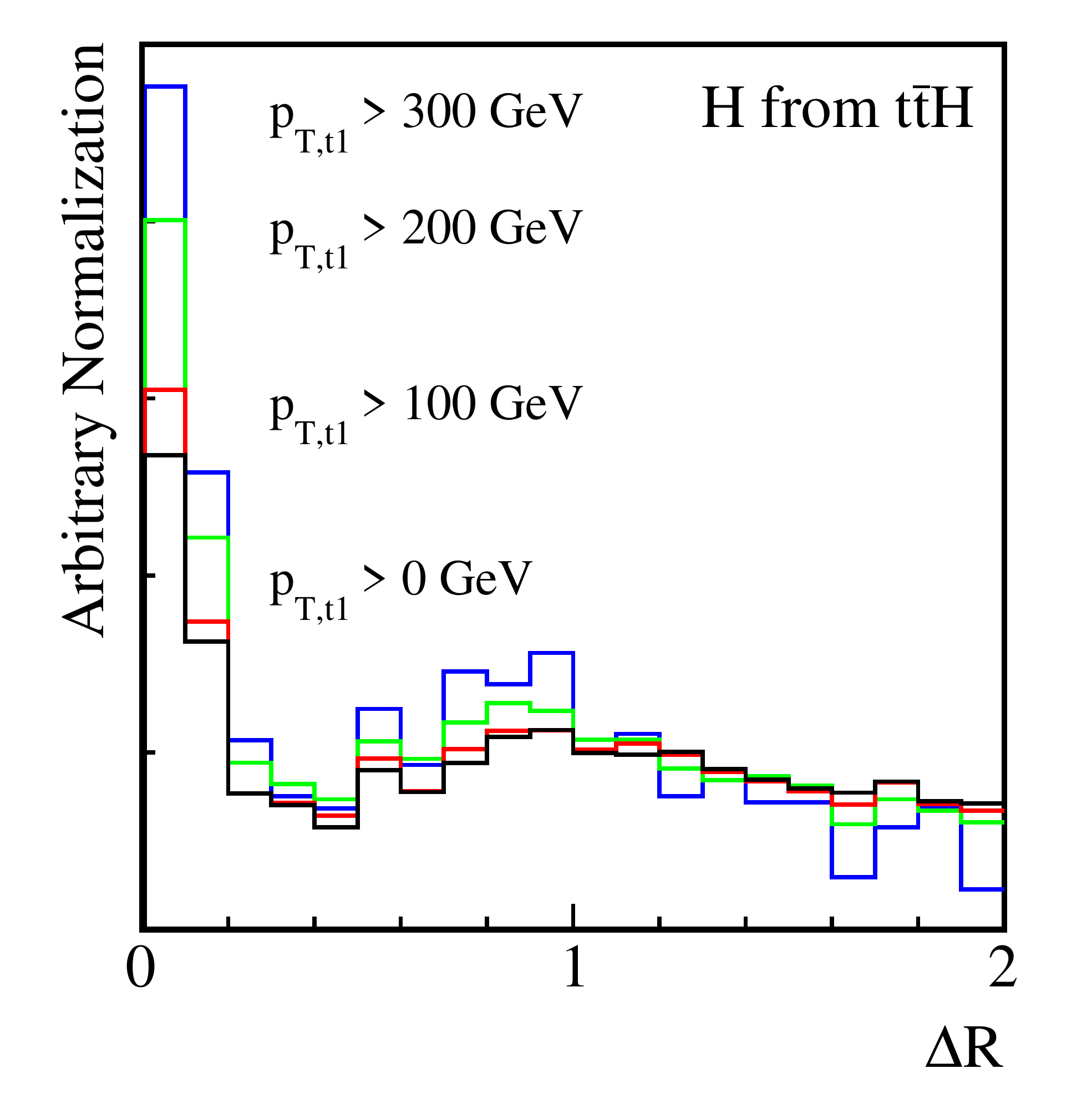} \\
\includegraphics[width=0.30\textwidth]{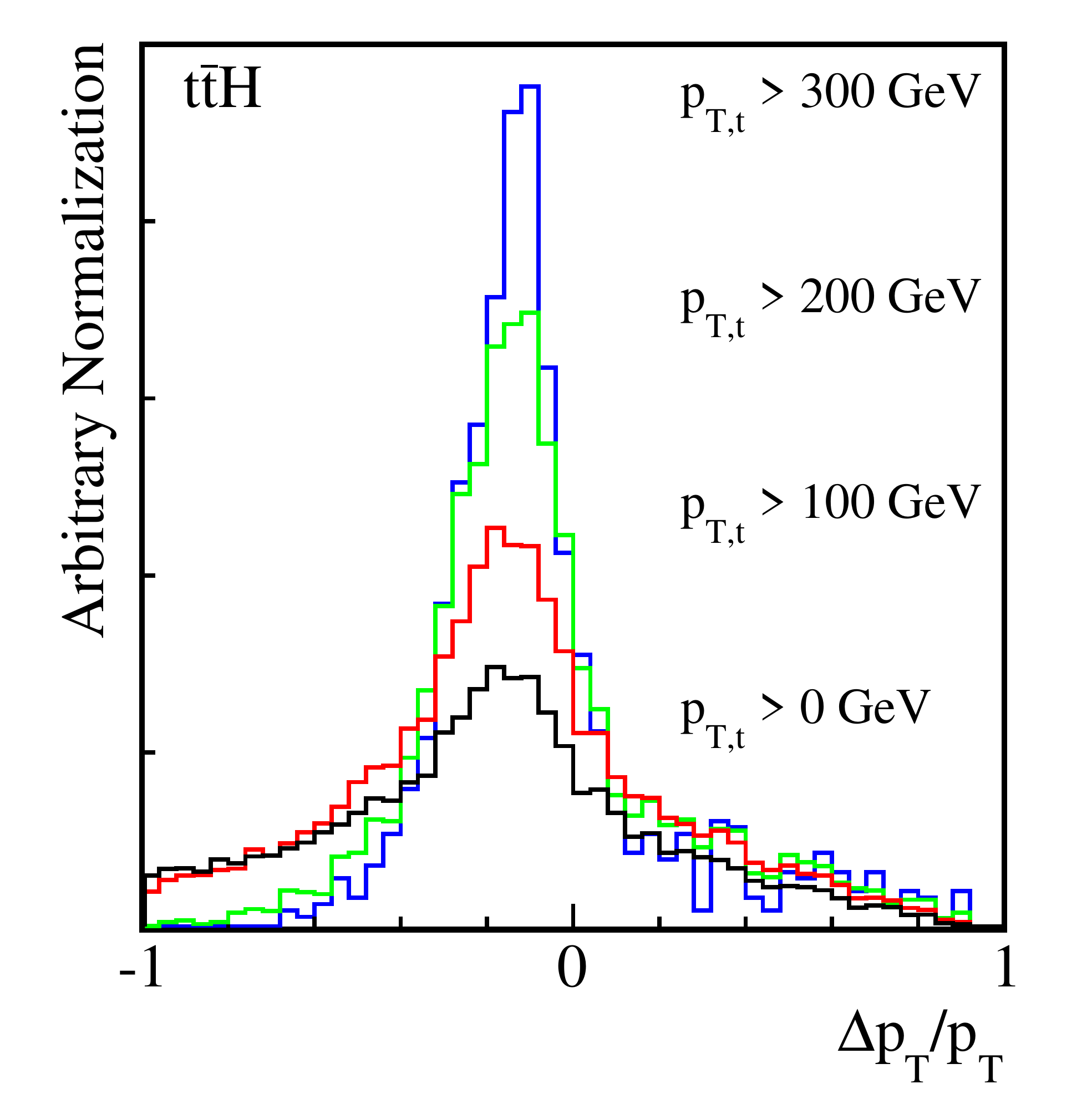}
\hspace*{0.02\textwidth}
\includegraphics[width=0.30\textwidth]{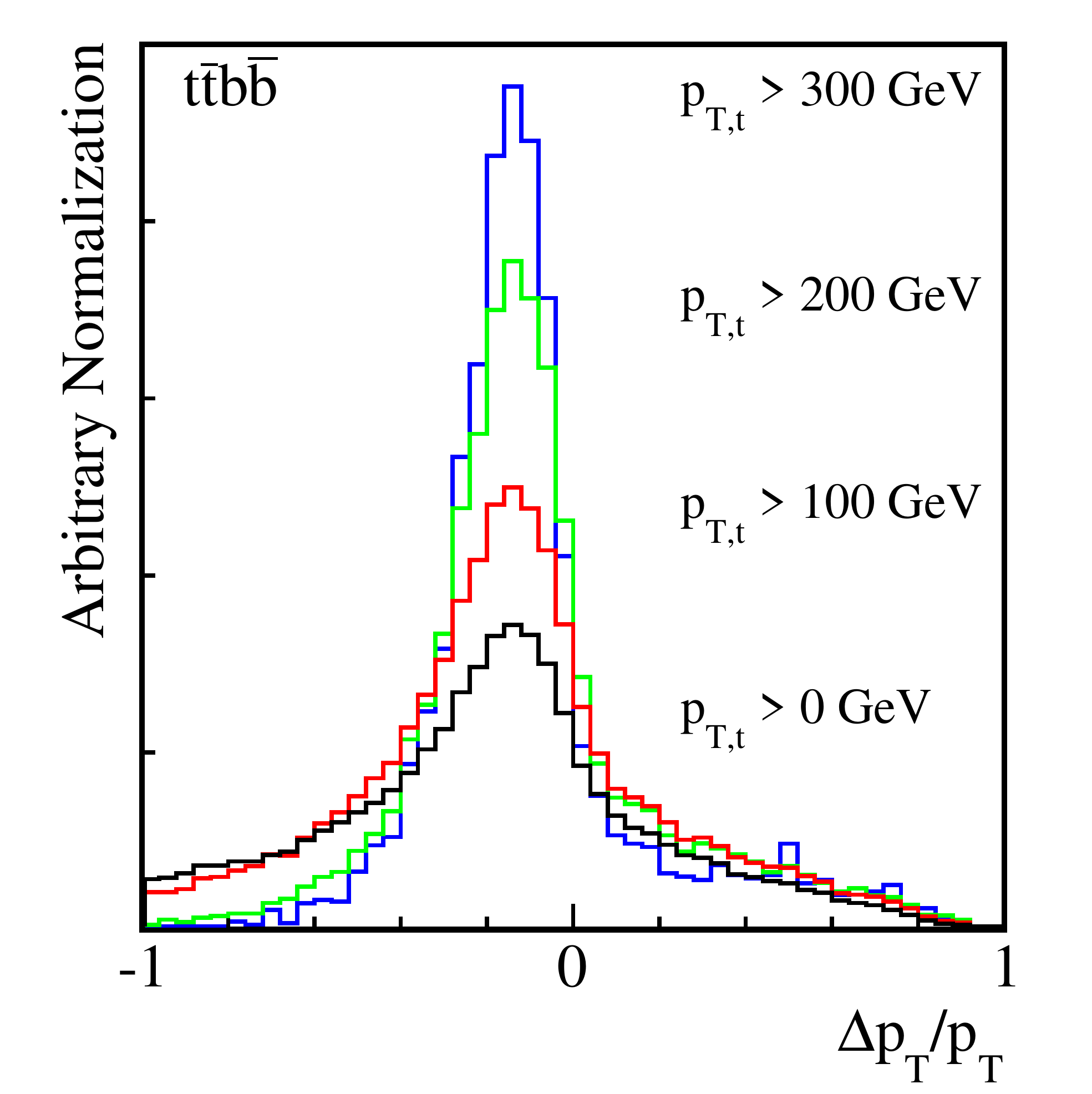}
\hspace*{0.02\textwidth}
\includegraphics[width=0.30\textwidth]{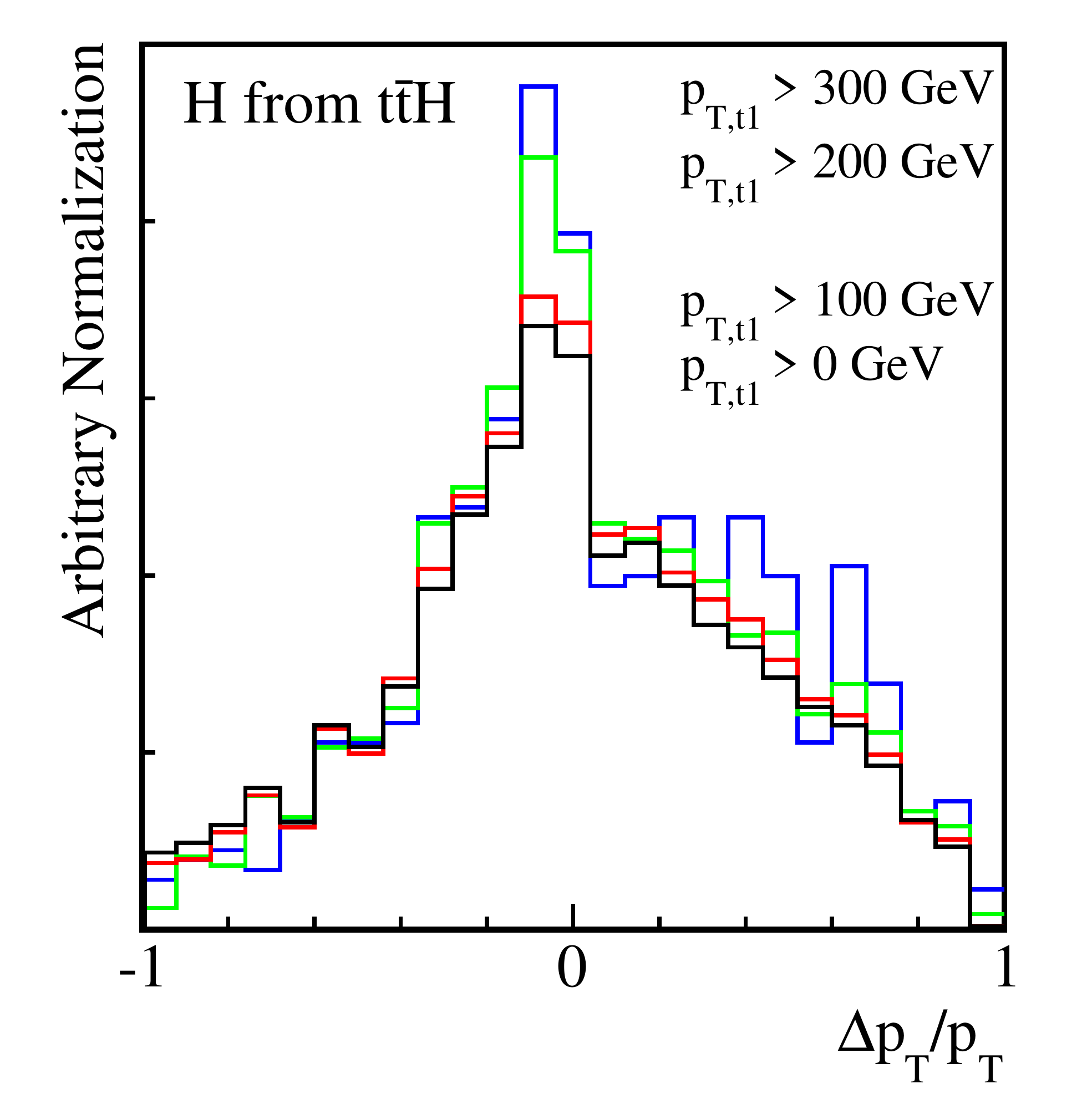}
\caption{$\Delta R$ (top row) and $\Delta p_T/p_T$ (bottom row) distributions 
for tops in $t\bar{t}H$ (left) and  $t\bar{t}b\bar{b}$ (center) samples, and 
for the Higgs in $t\bar{t}H$ (right). }
\label{fig:top_metric}
\end{figure}

We can compare the magnitude and direction of the reconstructed top
momenta to the true values of the parton--level tops, using Monte Carlo
truth. As in Ref.~\cite{buckets}, the two kinematic variables we
concentrate on are $\Delta R$, defined between the parton--level top or
Higgs and the closest of the two reconstructed tops or the
reconstructed Higgs in the event, and $\Delta p_T/ p_T^\text{bucket}$,
again taking the difference in $p_T$ between the parton--level object
and the nearest bucket-reconstructed top or the reconstructed Higgs,
normalized by the reconstructed $p_T$. Figure~\ref{fig:top_metric}
shows these distributions for both the signal and the irreducible
$t\bar{t}b\bar{b}$ background with real top quarks.  Different lines
correspond to the different reconstructed top $p_{T,t}^\text{min}$ cut
for each bucket.  For the Higgs plots (right column), these lines
correspond to the different $p_T$ cuts on the leading reconstructed
top in an event.

Defining a ``good'' reconstruction as $\Delta R < 0.5$ and $|\Delta
p_T /p_T |<0.2$ for the tops, we give in Table~\ref{tab:top_metric}
the percentage of reconstructed tops in the $t\bar{t}H$ signal and the
$t\bar{t}b\bar{b}$ background that are well-reconstructed. We also
show the percentage of well-reconstructed Higgses in the signal
sample. As can be seen, placing $p_T$ cuts on the reconstructed
tops increases the purity of well-reconstructed tops and Higgses, though
clearly this sacrifices total cross section.\bigskip

\begin{table}[b!]
\begin{tabular}{ll|rrrrrrrr}
\hline
&  & 0 GeV& 100 GeV& 150 GeV& 200 GeV& 250 GeV& 300 GeV \cr
\hline
$t$ from $t\bar{t}H$ 
& $\Delta R<0.5$& 0.357 & 0.515 & 0.643 & 0.759  & 0.820 & 0.856 \cr 
& $|\Delta p_T/p_T|<0.2$ & 0.256 & 0.378 & 0.452 & 0.518 & 0.558 & 0.586 \cr 
& $\Delta R<0.5$ and $|\Delta p_T/p_T|<0.2$ & 0.153 & 0.246 & 0.337 & 0.436 & 0.507 & 0.551 \cr 
\hline
$t$ from $t\bar{t}b\bar{b}$ 
& $\Delta R<0.5$  &0.415 & 0.563 & 0.681 & 0.777 & 0.837 & 0.860  \cr 
& $|\Delta p_T/p_T|<0.2$& 0.290 & 0.404 & 0.480 & 0.537 & 0.566 & 0.582 \cr 
& $\Delta R<0.5$ and $|\Delta p_T/p_T|<0.2$ &0.191 & 0.285 & 0.376 & 0.463  & 0.519 & 0.548  \cr 
\hline
$H$ from $t\bar{t}H$ 
& $\Delta R<0.5$ & 0.206 & 0.223  & 0.246 & 0.278 & 0.290 & 0.312  \cr 
& $|\Delta p_T/p_T|<0.2$& 0.290 & 0.301 & 0.319 & 0.330 & 0.331 & 0.325 \cr 
&$ \Delta R <0.5$ and $|\Delta p_T/p_T|<0.2$ & 0.116 & 0.128 & 0.143 & 0.162 & 0.172 & 0.189 \cr
\hline
\end{tabular}
\caption{Fraction of the buckets providing good momentum
  reconstruction for tops from $t\bar{t}H$ and $t\bar{t}b\bar{b}$, as
  well as Higgs from $t\bar{t}H$.  Percentages of well-reconstructed
  tops are shown after placing a $p_T$ cut on the top. For well-reconstructed Higgs,
  the percentages are shown for cuts on the highest-$p_T$ reconstructed top.}
\label{tab:top_metric}
\end{table}

\begin{figure}[t]
\includegraphics[width=0.30\textwidth]{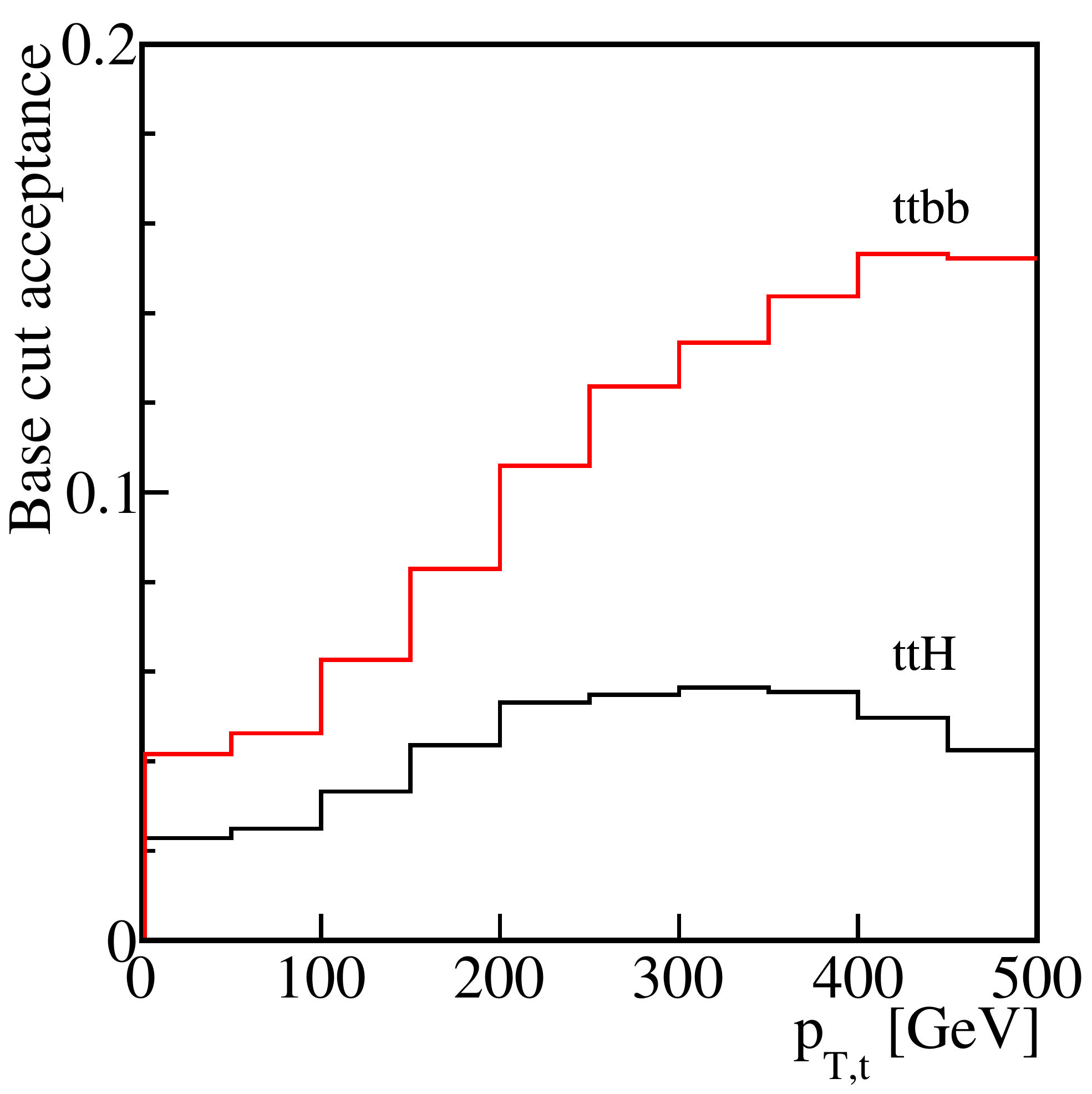}
\hspace*{0.02\textwidth}
\includegraphics[width=0.30\textwidth]{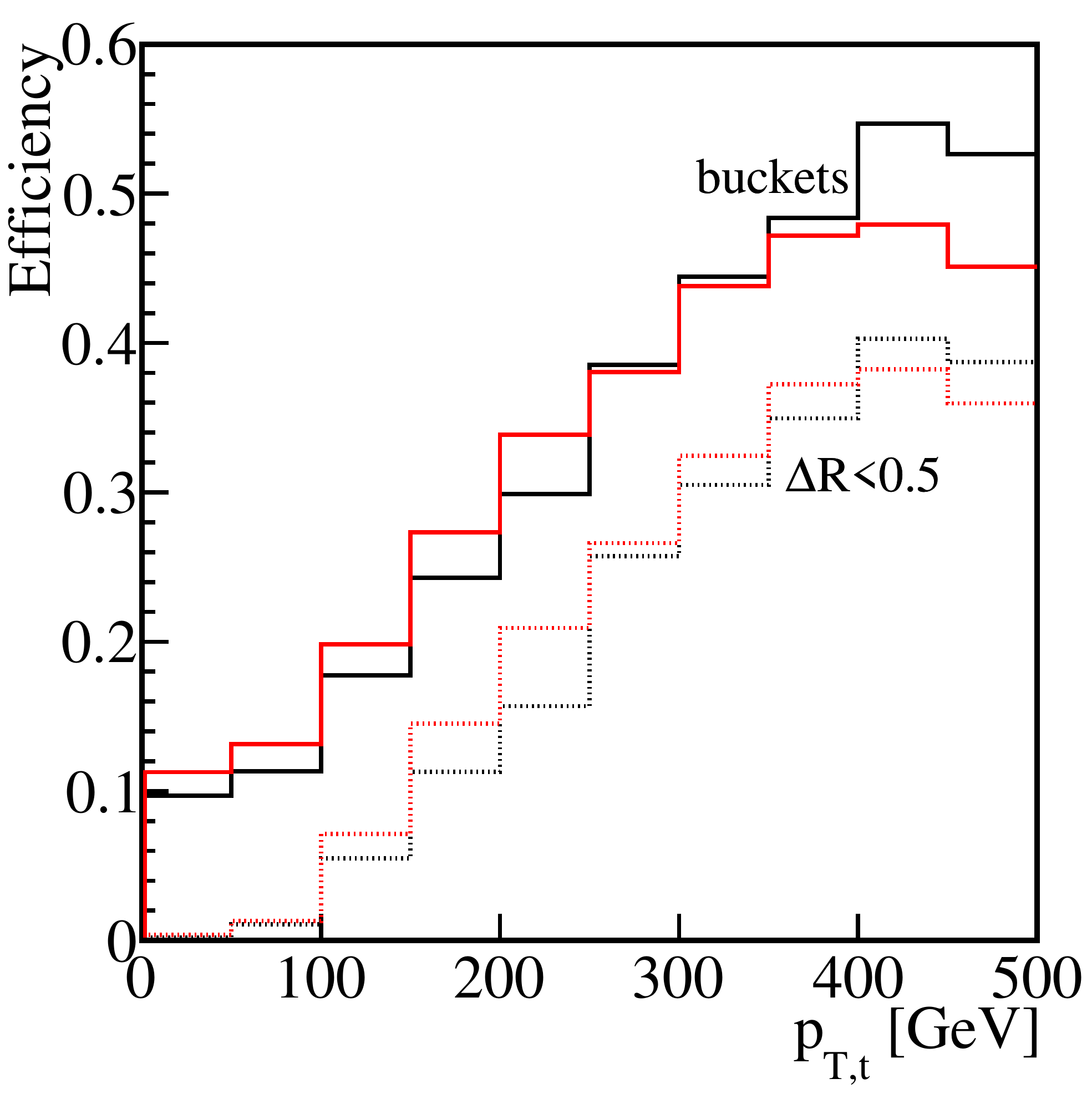}
\hspace*{0.02\textwidth}
\includegraphics[width=0.30\textwidth]{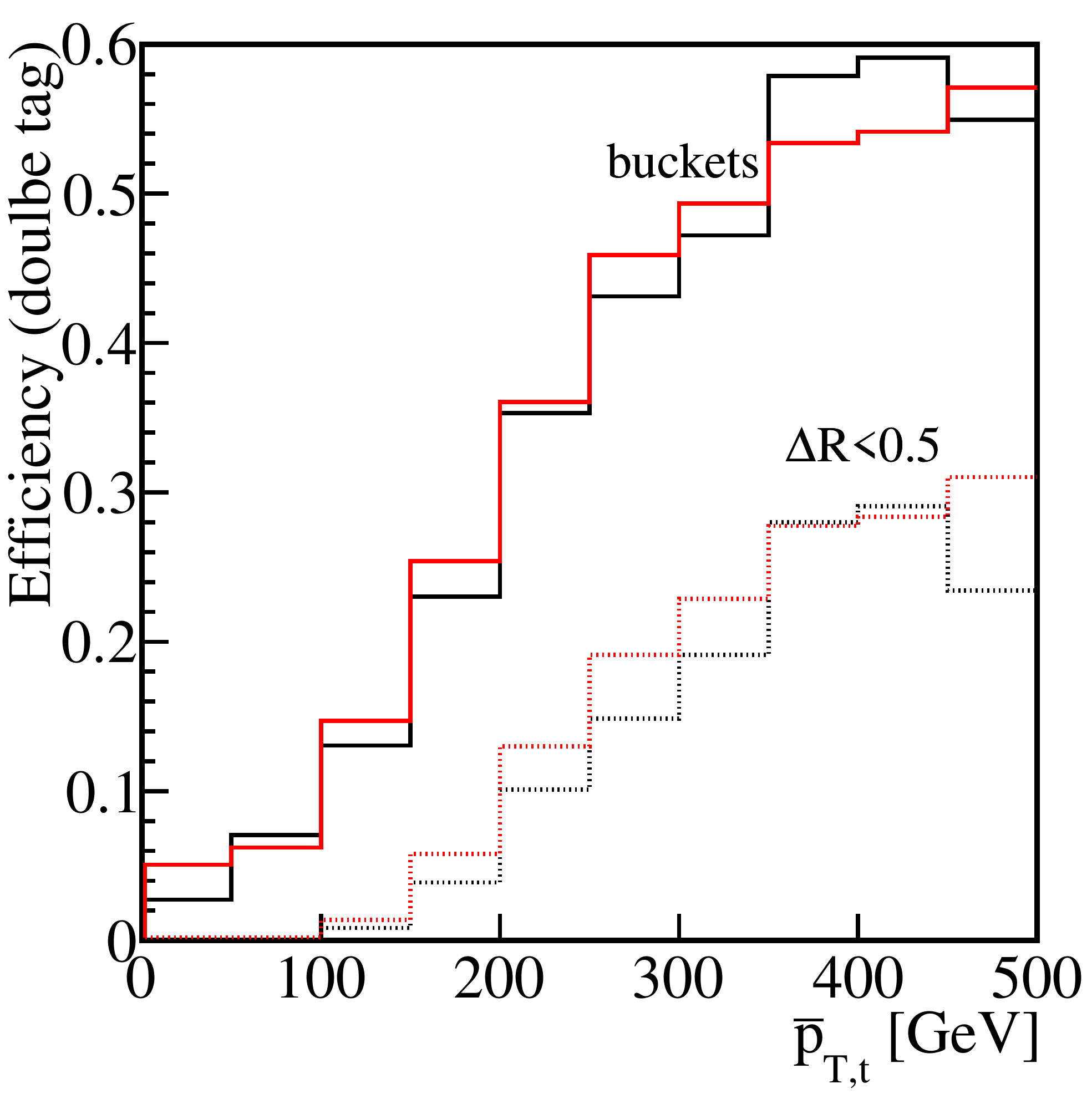}
\caption{Acceptance efficiency for the basic selection cut as a function of
  $p_{T,t}$ (left).  Efficiency for single/double bucket tag as a
  function of true transverse momentum of the top (center/right).}
\label{fig:recon_eff}
\end{figure}

Finally, in Figure~\ref{fig:recon_eff}, we show the efficiencies for this algorithm
 as a function of the parton--level top $p_T$,
for several ranges of reconstructed $p_T$. The left panel displays the
acceptances for the selection Eqs.\eqref{eq:global_cuts} and
\eqref{eq:global_jets} as a function of $p_{T,t}$. Note that the
acceptance for $t\bar{t}b\bar{b}$ sample is computed against the sample with the
generation cut of $p_{T,b}>35$~GeV and $R_{bb}>0.9$.  The central
panel shows the efficiency for a single reconstructed top as a
function of the parton--level top $p_T$ relative to the number in the
left panel. The efficiencies for $t\bar{t}H$ (black) and $t\bar{t}b\bar{b}$ (red) are
shown. The contributions for the buckets with a parton--level top
found in $\Delta R<0.5$ are indicated by dotted lines. We see both
channels have similar efficiencies after the acceptance cut. The right
panel gives the double tag efficiency as a function of the average of
the parton--level transverse momenta of the two tops. Note that our
algorithm always tags two tops and the resulting efficiencies are
similar to the central panel in number.

\end{document}